\documentclass{aa}
\usepackage{natbib}
\usepackage{graphicx}

\newcommand\dnu{\Delta\nu}
\newcommand\dnumoy{\langle\Delta\nu\rangle}
\newcommand\numax{\nu\ind{max}}
\newcommand\nuc{\nu\ind{c}}

\newcommand\nuenv{\delta\nu\ind{env}}
\newcommand\hauteur{\mathcal{H}}
\newcommand\background{\mathcal{B}}
\newcommand\HBR{\hauteur/\background}

\newcommand\iref{_\odot}
\newcommand\dnus{\dnu\iref}
\newcommand\numaxs{\numax{}_{,}{}\iref}
\newcommand\Ts{T\iref}
\newcommand\Rs{R\iref}
\newcommand\Ms{M\iref}
\newcommand\Rsis{R} 
\newcommand\Msis{M} 

\newcommand\deltanunu{\Delta\nu(\nu)}

\newcommand\ampmax{\mathcal{F}}
\newcommand\Abol{A\ind{max}}

\newcommand\nombre{\mathcal{N}}
\newcommand\nzero{\nombre_0}\newcommand\nun{\nombre_1}\newcommand\nde{\nombre_2}\newcommand\ntr{\nombre_3}
\newcommand\npeak{\bar{N}\ind{peak}}
\newcommand{\ind}[1]{_{\mathrm{#1}}}

\def\muHz{\,$\mu$Hz}

\def\m2s2{\,m$^{2}$\,s$^{-2}$} 

\def\Teff{T\ind{eff}}

\def\deltanunu{\Delta\nu(\nu)}

\begin{document}
\title{
Red-giant seismic properties analyzed with CoRoT\thanks{The CoRoT space mission, launched on 2006 December 27, was developed and is operated by the CNES, with participation of the Science Programs of ESA, ESA's RSSD, Austria, Belgium, Brazil, Germany and Spain.}}
\titlerunning{CoRoT red giants}
\author{
B. Mosser\inst{1}\and
K. Belkacem\inst{2,1}\and
M.-J. Goupil\inst{1}\and
A. Miglio\inst{2,}\thanks{Postdoctoral Researcher, Fonds de la Recherche Scientifique - FNRS, Belgium}\and
T. Morel\inst{2}\and
C. Barban\inst{1}\and
F. Baudin\inst{3}\and
S. Hekker\inst{4,5}\and
R. Samadi\inst{1}\and
J. De Ridder\inst{5}\and
W. Weiss\inst{6}\and
M. Auvergne\inst{1}\and
A. Baglin\inst{1}
}

\offprints{B. Mosser}

\institute{LESIA, CNRS, Universit\'e Pierre et Marie Curie, Universit\'e Denis Diderot, Observatoire de Paris, 92195 Meudon cedex, France\\
\email{benoit.mosser@obspm.fr}
\and
Institut d'Astrophysique et de G\'eophysique, Universit\'e de Li\`ege, All\'ee du 6 Ao\^ut, 17 B-4000 Li\`ege, Belgium
\and
Institut d'Astrophysique Spatiale, UMR 8617, Universit\'e Paris XI, B\^atiment 121, 91405 Orsay Cedex, France
\and
School of Physics and Astronomy, University of Birmingham, Edgbaston, Birmingham B15 2TT United Kingdom
\and
Instituut voor Sterrenkunde, K. U. Leuven, Celestijnenlaan 200D, 3001 Leuven, Belgium
\and
Institut f\"ur Astronomie (IfA), Universit\"at Wien, T\"urkenschanzstrasse 17, 1180 Wien, Austria
}

\date{Accepted in A\&A}

\abstract{The CoRoT 5-month long observation runs give us the opportunity to analyze a large variety of red-giant stars and to derive fundamental stellar parameters from their asteroseismic properties.}%
{We perform an analysis of more than 4\,600 CoRoT light curves to extract as much information as possible. We take into account the characteristics of the star sample and of the method in order to provide asteroseismic results as unbiased as possible. We also study and compare the properties of red giants of two opposite regions of the Galaxy.}%
{
We analyze the time series with the envelope autocorrelation function in order to extract precise asteroseismic parameters with reliable error bars. We examine first the mean large frequency separation of solar-like oscillations and the frequency of maximum seismic amplitude, then the parameters of the excess power envelope. With the additional information of the effective temperature, we derive the stellar mass and radius.}%
{We have identified more than 1\,800 red giants among the 4\,600 light curves and have obtained accurate distributions of the stellar parameters for about 930 targets. We were able to reliably measure the mass and radius of several hundred red giants. We have derived precise information on the stellar population distribution and on the red clump. Comparison between the stars observed in two different fields shows that the stellar asteroseismic properties are globally similar, but with different characteristics for red-clump stars.}%
{This study shows the efficiency of statistical asteroseismology: validating scaling relations allows us to infer fundamental stellar parameters, to derive precise information on the red-giant evolution and interior structure and to analyze and compare stellar populations from different fields.}

\keywords{stars: fundamental parameters  -- stars: interiors -- stars: evolution -- stars: oscillations -- stars: red giants}
\maketitle
\section{Introduction}

The high-precision, continuous, long  photometric time series recorded by the CoRoT satellite allow us to study a large number of red giants.
In a first analysis of CoRoT red giants, \cite{2009Natur.459..398D} reported the presence of radial and non-radial oscillations in more than 300 giants.
\cite{2009A&A...506..465H}, after a careful analysis of about 1000 time series, have demonstrated a tight relation between the large separation and the frequency of maximum oscillation amplitude.  \cite{2009A&A...503L..21M} have identified the signature of the red clump, in agreement with   synthetic populations.
\cite{2009ASPC..404..307K} have exploited the possibility of measuring stellar mass and radius from the asteroseismic measurements, even when the stellar luminosity and effective temperature are not accurately known.

In this paper, we specifically focus on the statistical analysis of a large set of stars in two different fields observed with CoRoT \citep{2009A&A...506..411A}. One is located towards the Galactic center (LRc01), the other in the opposite direction  (LRa01). We first derive precise asteroseismic parameters, and then stellar parameters. We also examine how these parameters vary with the  frequency $\numax$ of the maximum amplitude. The new analysis that we present in this paper has been made possible by the use of the autocorrelation method \citep{2009A&A...508..877M}, which is significatively different from what has been used in other works \citep{2010A&A...511A..46M, 2009A&A...506..465H, 2009CoAst.160...74H}. It does not rely on the identification of the excess oscillation power, but on the direct measurement of the acoustic radius $\tau$ of a star. This acoustic radius is  related to the large separation commonly used in asteroseismology ($\dnu = 1/2\tau$).
The chronometer is provided by the autocorrelation of the asteroseismic time series, which is sensitive to the travel time of a pressure wave crossing the stellar diameter twice, i.e. 4 times the acoustic radius.
The calculation of this autocorrelation as the Fourier spectrum of the Fourier spectrum with the use of narrow window for a local analysis in frequency has been suggested by \cite{2006MNRAS.369.1491R}. \cite{2009A&A...508..877M} have formalized and quantified the performance of the method based on the envelope autocorrelation function (EACF).

With this method, and the subsequent automated pipeline, we search for the signature of the mean large separation of a solar-like oscillating signal in the autocorrelation of the time series. \cite{2009A&A...508..877M} have shown how to deal with the noise contribution entering the autocorrelation function, so that they were able to determine the reliability of the large separations obtained with this method. Basically, they have scaled the autocorrelation function according to the noise contribution. With this scaling, they have shown how to define the threshold level above which solar-like oscillations are detected and a reliable large separation can be derived.

An appreciable advantage of the method consists in the fact that the large separation is determined first, without any assumptions or any fit of the background. As a consequence, the method directly focusses on the key parameters of asteroseismic observations: the mean value $\dnumoy$ of the large separation and the frequency $\numax$ where the oscillation signal is maximum. Since it does not rely on the detection of an energy excess, it can operate at low signal-to-noise ratio, as shown by \cite{2009A&A...506...33M}. The value of the frequency $\numax$, inferred first from the maximum autocorrelation signal, is then refined from the maximum excess power observed  in a smoothed Fourier spectrum corrected from the background component. The different steps of the pipeline for the automated analysis of the time series are presented in \citet{ma10}.

The method has been tested on CoRoT main-sequence stars \citep{2009A&A...507L..13B, 2009A&A...506...51B, 2009A&A...506...41G, Ballot2010, Deheuvels2010, Mathur2010} and has proven its ability to derive reliable results rapidly for low signal-to-noise ratio light curves, when other methods fail or give questionable results \citep{2009A&A...506...33M, Gaulme2010}.
The method also allowed the correct identification of the degree of the eigenmodes of the first CoRoT target HD49933 \citep{2005A&A...431L..13M, 2008A&A...488..705A, 2009A&A...508..877M}. The EACF method and its automated pipeline have been tested on the CoRoT red giants presented by  \cite{2009Natur.459..398D}  and \cite{2009A&A...506..465H}, and also on the Kepler red giants \citep{Stello2010, Bedding2010}.

The paper is organized as follows. In Sect.~\ref{methode}, we present the analysis of the CoRoT red giants using the EACF and define the way the various seismic parameters are derived. We also determine the frequency interval where we can extract unbiased global information. Measurements of the asteroseismic parameters $\dnumoy$ and $\numax$ are presented in Sect.~\ref{scaling} and compared to previous studies. We also present the variation $\deltanunu$ performed with the EACF.
Section~\ref{power} deals with the parameters related to the envelope of the excess power observed in the Fourier spectra, for which we propose scaling laws.
From the asteroseismic parameters $\numax$ and $\dnumoy$, we determine the red-giant mass and radius in Sect.~\ref{masserayon}. Compared to \cite{2009ASPC..404..307K}, we benefit from the stellar effective temperatures obtained from independent photometric measurements, so that we do not need to refer to stellar modeling for deriving the fundamental parameters. We then specifically address the properties of the red clump in Sect.~\ref{redclump}, so that we can carry out a quantitative comparison with the synthetic population performed by \cite{2009A&A...503L..21M}. The difference between the red-giant populations observed in 2 different fields of view is also presented in Sect.~\ref{redclump}. Section~\ref{conclusion} is devoted to discussions and conclusions.


\begin{table}
\caption{Red-giant targets}\label{runs}
\begin{tabular}{lccccc}
\hline
run  & $T$ (d)&$\nzero$& $\nun$ & $\nde$& $\ntr$ \\
\hline
LRc01 & 142 d& 9938 & 3388   & 1399    & 710    \\
LRa01 & 128 d& 2826 & 1271   &  428    & 219    \\
\hline
total&  --  &12764 & 4659   & 1827    & 929    \\
\hline
\multicolumn{2}{l}{ratio (\%)} &  --   &100\,\% & 39\,\%  & 20\,\% \\
\hline
\end{tabular}

Among $\nzero$ targets a priori identified as red giants in the input catalog of each field, $\nun$ light curves were available and analyzed in each run. Among these, $\nde$ targets show solar-like oscillation patterns for which we can derive precise values of $\dnumoy$ and $\numax$. Envelope parameters can be precisely determined for $\ntr$ targets.
\end{table}

\section{Data\label{methode}}

\subsection{Time series\label{obs}}

Results are based on time series recorded during the first long CoRoT runs in the direction of the Galactic center (LRc01) and in the opposite direction (LRa01). These long runs last approximately 140 days, providing us with a frequency resolution of about 0.08\muHz. Red giants were identified according to their location in a colour-magnitude diagram with $J-K$ in the range $[0.6, 1.0]$ and $K$ brighter than 12.

We present in Table \ref{runs} the number of targets that were considered. $\nzero$ is the number of red giants identified in each field according to a colour-magnitude criterion. Only a fraction of these targets were effectively observed. $\nun$ then represents the number of time series available, hence analyzed.
Among the $\nun$ time series,  $\nde$ targets show reliable solar-like oscillations for which we can derive precise values of $\dnumoy$ and $\numax$. We remark that the ratio $\nde/\nun$ is high: a large fraction of the stars identified as red-giant candidates presents solar-like oscillations.

\subsection{Data analysis}

As explained by \citet{ma10}, the measurement of the mean value $\dnumoy$ of the large separation presupposes scaling relations between this parameter, the frequency of maximum seismic amplitude $\numax$ and the full-width at half-maximum of the excess power envelope $\nuenv$. These scaling relations are used  to search for the optimized asteroseismic signature. The threshold level for a positive detection of solar-like oscillations and the quality of the signature are given by the maximum $\ampmax$ reached by the EACF \citep[paragraph 3.3 of][]{2009A&A...508..877M}. The method is able to automatically exclude unreliable results and calculate error bars \emph{without any comparison to theoretical models}.

For stars with low signal-to-noise seismic time series, only $\dnumoy$ and $\numax$ can be reliably estimated. At larger signal-to-noise ratio, we can also derive the parameters of the envelope corresponding to the oscillation energy excess. This envelope is supposed to be Gaussian, centered on $\numax$, with a full-width at half-maximum $\nuenv$. We also measure the height-to-background ratio $\HBR$ in the power spectral density smoothed with a $\dnumoy$-broad cosine filter as the ratio between the excess power height $\hauteur$ and the activity background $\background$. The determination of these envelope parameters requires in fact a high enough height-to-background ratio ($\ge 0.2$).  Finally, the maximum amplitude of the modes and the FWHM of the envelope were precisely determined for $\ntr$ targets, for which precise measurements of the stellar mass and radius can then be derived.

\begin{figure}
\centering
\includegraphics[width=8.5cm]{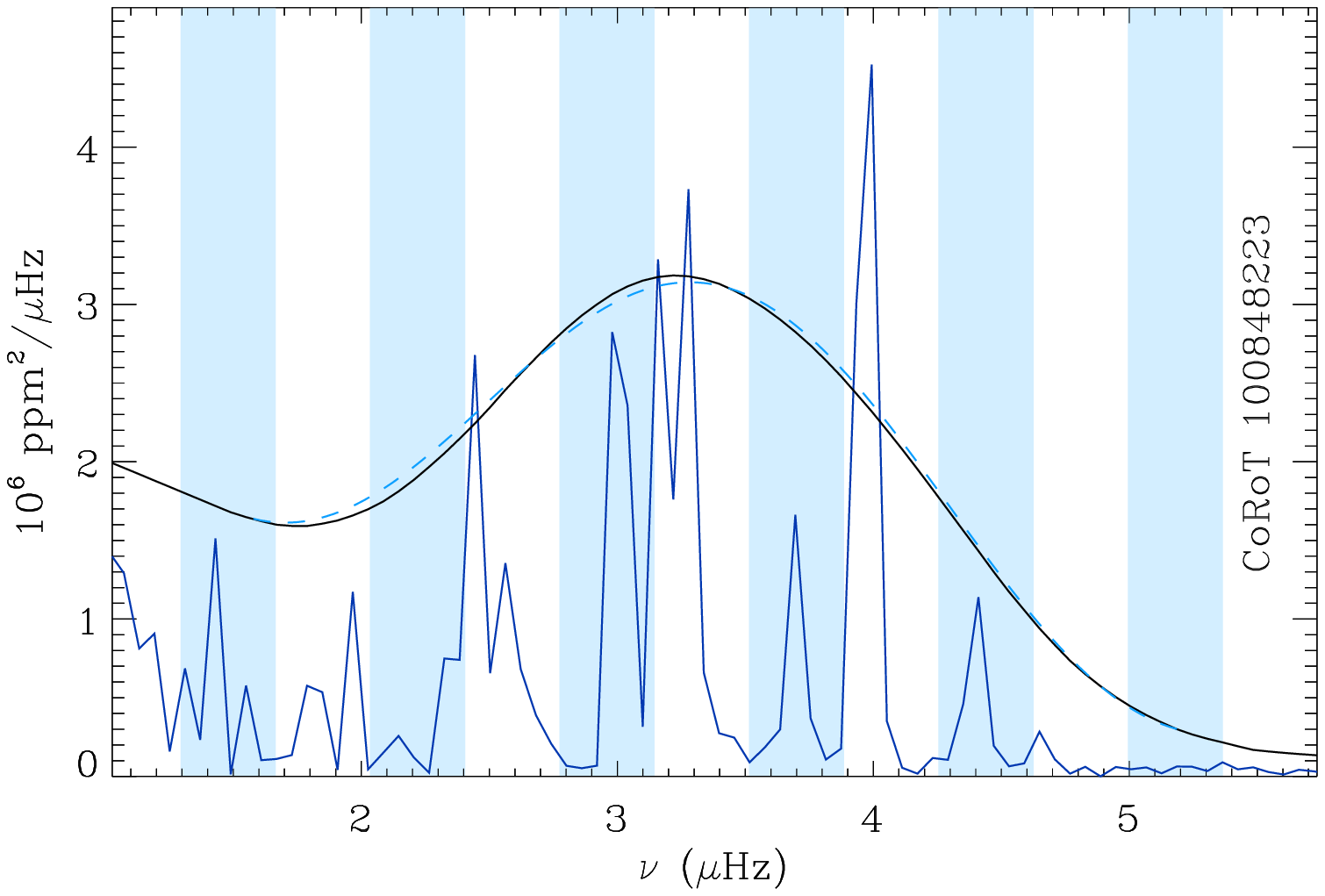}
\caption{Fourier spectrum of a target with a very low mean value of the large separation ($\dnumoy=0.74$\muHz , centered at $\numax = 3.45$\muHz ). The colored vertical ranges have a width equal to half the large separation.
This spectrum exhibits a clear Tassoul-like pattern: modes of degree 0 and 2 are located in the uncolored regions whereas $\ell=1$ modes are in the blue regions. For clarity, the amplitude of the envelope (black line) has been multiplied by 3. The dashed line represents the Gaussian fit of the excess power envelope, also multiplied by 3, superimposed on the background.
\label{spectres}}
\end{figure}

Thanks to the length of the runs and to the long-term stability of CoRoT, large separations below 1\muHz\ have been measured for the first time. This represents about 10 times the frequency resolution of 0.08\muHz. We want to stress that the method based on the EACF allows us to obtain a better resolution since the achieved precision is related to the ratio of the time series sampling to the acoustic radius \citep[see Eq.\,A.8 of][]{2009A&A...508..877M}. We can reach a frequency resolution of about 3\,\% when the excess power envelope is reduced to 3 times the large separation. Figure~\ref{spectres} gives an example of the fits obtained at low frequency. The CoRoT star 100848223 has a mean large separation $\dnumoy=0.74\pm0.02$\muHz\ and a maximum oscillation frequency $\numax=3.45\pm0.22$\muHz.

\begin{figure}
\centering
\includegraphics[width=8.5cm]{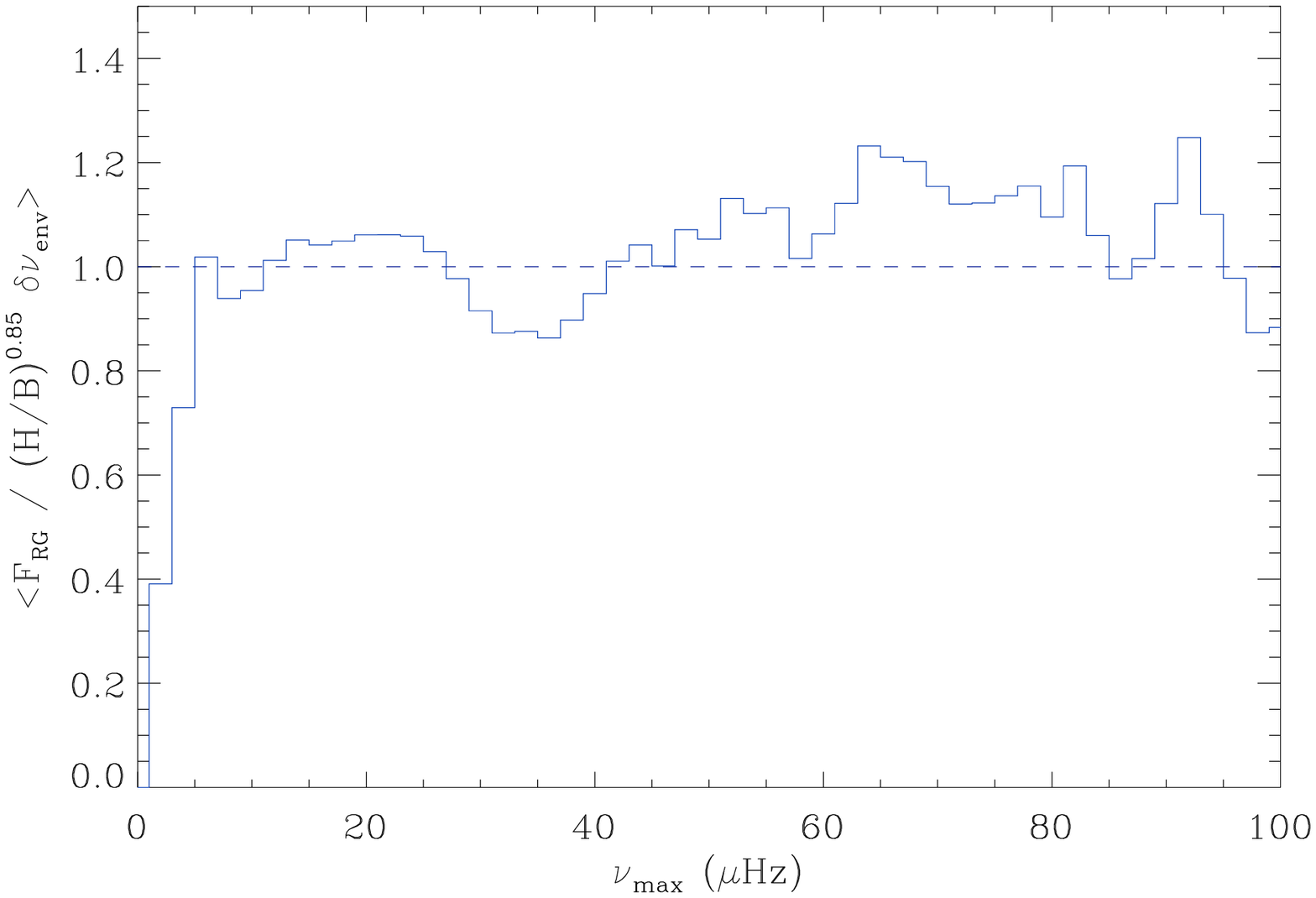}
\caption{Estimation of the bias, calculated from the mean ratio $ {\ampmax}\ind{RG} / (\HBR)^{0.85}\, \nuenv$ as a function of the frequency of maximum amplitude $\numax$.
\label{deterbiais}}
\end{figure}

\begin{figure*}
\centering
\includegraphics[width=13.5cm]{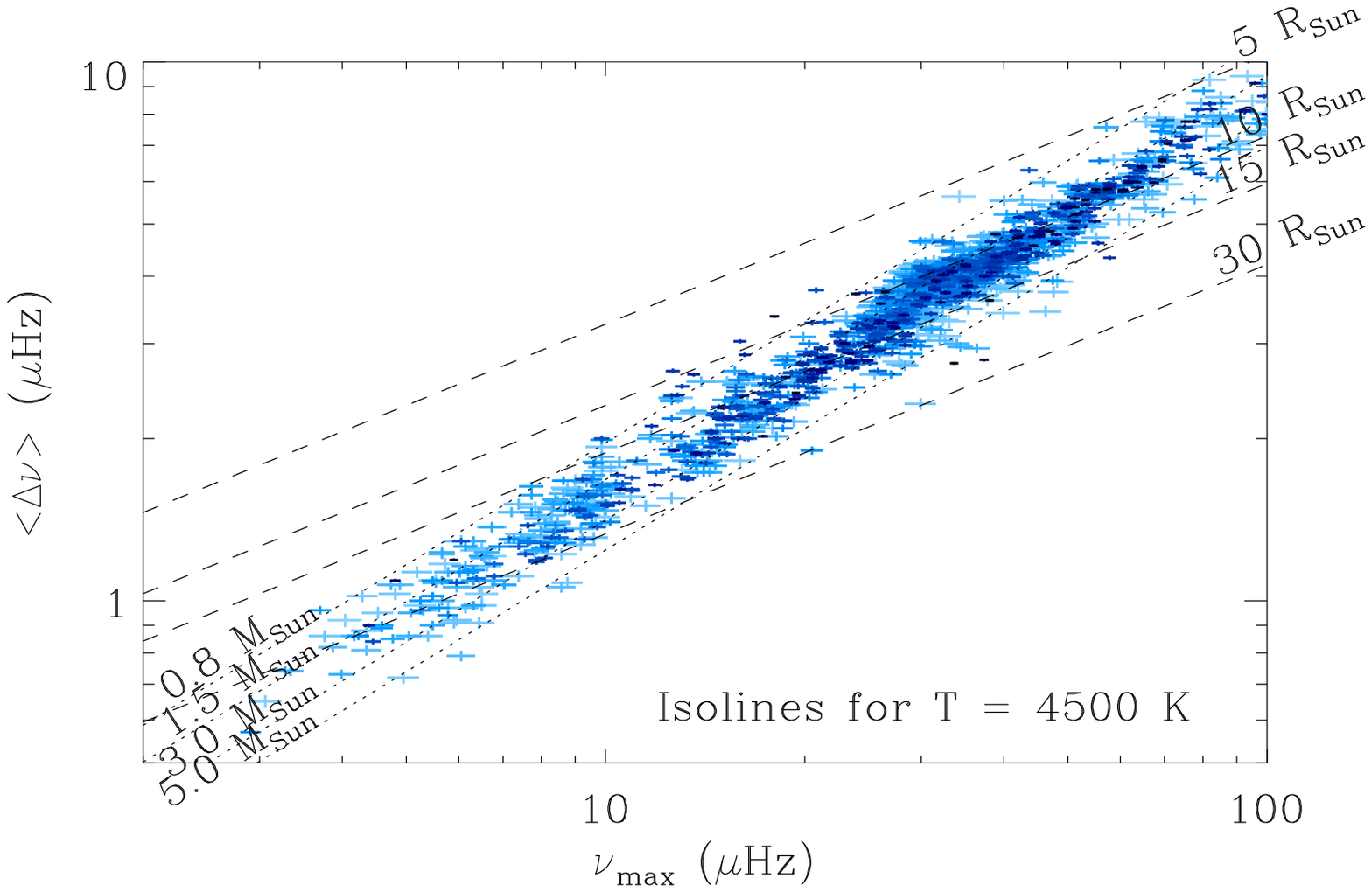}
\caption{$\numax$ - $\dnumoy$ relation for red-giant stars in LRa01 and LRc01, with all data satisfying a rejection of the null hypothesis at the 1\,\% level. Isoradius and isomass lines, derived from the scaling given by Eq.~\ref{rayon} and \ref{masse}, are given for a mean effective temperature of 4\,500\,K. Error bars in $\numax$ and $\dnumoy$ are derived from the amplitude $\ampmax$ of the EACF. The colour code allocates a darker colour for measurements with high $\ampmax$.
\label{numax_dnu}}
\end{figure*}

\subsection{Bias and error bars}

The distribution of the targets can be biased by different effects that have to be carefully examined before extracting any statistical properties. Since we aim to relate global properties to $\numax$, we have examined how the distribution of red giants can be biased as a function of this frequency.
We chose to consider only targets with $\numax$ below 100\muHz. For values above that level, the oscillation pattern can be severely affected by the orbit, at frequencies mixing the orbital and diurnal signatures ($161.7\pm k\,11.6$\muHz, with $k$ an integer). This high-frequency domain will be more easily  studied with Kepler  \citep{Bedding2010}.

On the other hand, bright stars with a large radius, hence a low mean density, present an oscillation pattern at very low frequency. In that respect, even if CoRoT has provided the longest continuous runs ever observed, these bright targets that should be favored by a magnitude effect suffer from the finite extent of the time series. The EACF allows us to examine the bias in the data, via the distribution of the autocorrelation signal as a function of frequency.

According to \cite{2009A&A...508..877M}, the EACF amplitude  $\ampmax$ scales as $(\HBR)^{1.5} \nuenv$. This factor $\ampmax$ measures the quality of the data, since the relative precision of the measurement of $\dnumoy$ and $\numax$ varies as $\ampmax^{-1}$.
Contrary to the linear dependence with $\nuenv$, which has been theoretically justified by \cite{2009A&A...508..877M}, the variation of $\ampmax$ with $\HBR$ was empirically derived from a fit based on main-sequence stars. We have verified that this relation for the variation of $\ampmax$ with $\HBR$ \emph{cannot} be extrapolated to red giants. The reason seems to be related to the difference between the oscillation patterns of red giants compared to main-sequence stars \citep{2009A&A...506...57D}.
For $\numax \le 80$\muHz, the number of targets showing solar-like oscillation is high enough to derive the exponent for giants, close to 0.85:
\begin{equation}\label{precision}
   {\ampmax}\ind{RG} \propto \left( {\hauteur\over\background} \right)^{0.85} \ \nuenv .
\end{equation}

Due to the very large number of red giants and to the large variety of the targets, the distribution of the ratio ${\ampmax}\ind{RG}/ (\HBR)^{0.85} \,\nuenv$ is broadened compared to the few solar-like stars considered in \cite{2009A&A...508..877M}. Its mean value shows a clear decrease at frequencies below 6\muHz\ and a plateau at higher frequency (Fig.~\ref{deterbiais}). This can be interpreted as a deficit of high signal-to-noise data when $\numax<6$\muHz, hence a signature of a bias against low $\numax$ values. Above 80\muHz, the number of targets is small and these targets present high EACF but poor $\HBR$ ratio; however, the fit presented in Eq.~(\ref{precision}) remains valid. The observed decrease of the number of targets with increasing $\numax$ is however coherent with the extrapolation from lower values, with the observations of \cite{2009A&A...506..465H} and with Kepler data \citep{Bedding2010}.

We conclude from this test that the distribution of the targets is satisfactorily sampled in the frequency range [3.5, 100\muHz], with no bias introduced by the method above 6\muHz\ and especially in the most-populated area in the range [30, 40\muHz] corresponding to the red-clump stars \citep{1999MNRAS.308..818G, 2009A&A...503L..21M}.

\begin{figure}
\centering
\includegraphics[width=8.5cm]{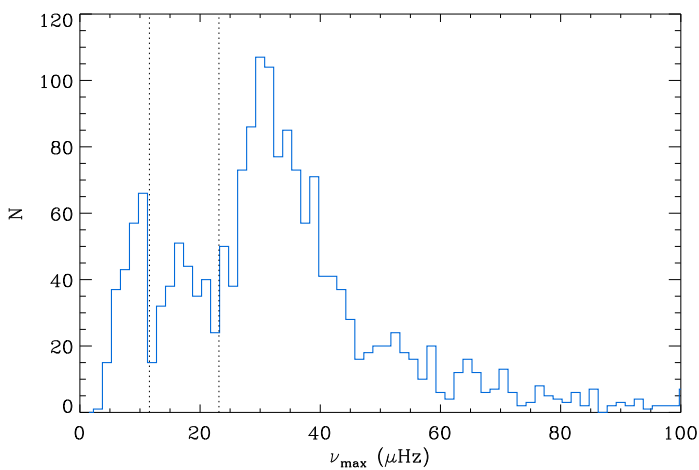}
\caption{Histogram of $\numax$ with the signature of the red clump in the range [30, 40\muHz], with a major contribution at 30 and a shoulder plus a secondary bump just below 40\muHz. The deficits around 11.6 and 23.2\muHz\ (vertical dotted lines) are artifacts due the low-Earth orbit of CoRoT.
\label{histo_numax}}
\end{figure}

\begin{figure}
\centering
\includegraphics[width=8.5cm]{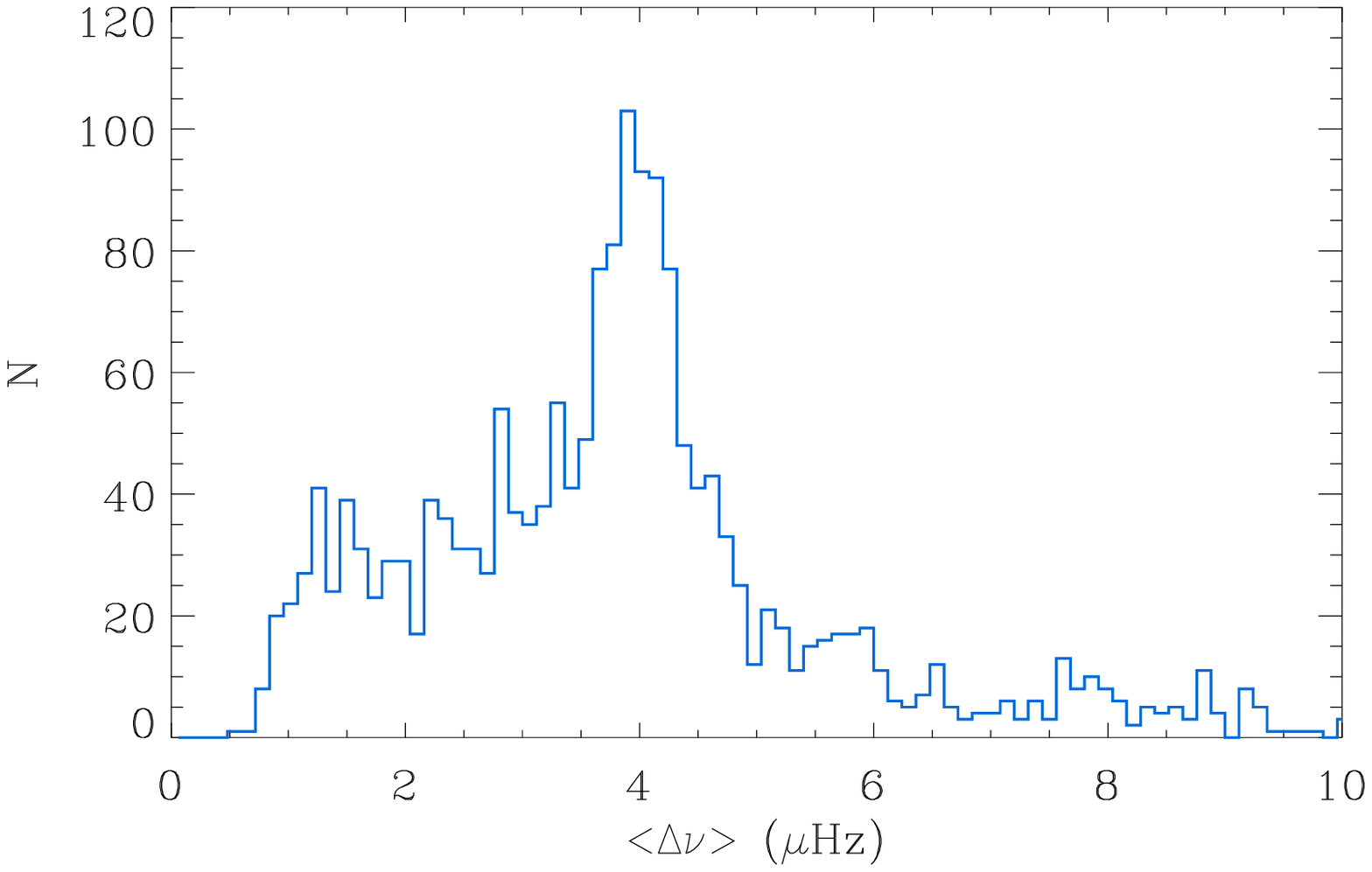}
\caption{Histogram of $\dnumoy$, with the red clump signature around 4\muHz.
\label{histo_dnu}}
\end{figure}

\section{Frequency properties\label{scaling}}

\subsection{Mean large separation and frequency of maximum amplitude\label{dnu}}

The mean large separation and the frequency of maximum amplitude have the most precise determination. The median values of the 1-$\sigma$ uncertainties on $\dnumoy$ and $\numax$ are respectively about 0.6 and 2.4\,\%. The scaling between $\numax$ and $\dnumoy$ reported by \cite{2009A&A...506..465H} for red giants and discussed by \cite{2009MNRAS.400L..80S} has been explored down to $\numax= 3.5$\muHz\ (or $\dnumoy= 0.75$\muHz). We obtain a more precise determination of the scaling (Fig.~\ref{numax_dnu}), with more than 1\,300 points entering the fit:
\begin{equation}\label{eqt__scaling}
   \dnumoy \simeq (0.280 \pm 0.008)\ \numax^{0.747\pm0.003}
\end{equation}
with $\dnumoy$ and $\numax$ in \muHz. The 1-$\sigma$ errors given in Eq.~(\ref{eqt__scaling}) are internal errors and cannot be considered as significant.
We note that a modification of the data sample, for instance by reducing the frequency interval or the number of data, yields variations greater than the internal error bars. Hence, a more realistic relation with conservative error bars is:
\begin{equation}\label{eqt_scaling}
   \dnumoy \simeq (0.280 \pm 0.02)\ \numax^{0.75\pm0.01}
\end{equation}
with error bars that encompass the dispersion in the different sub-samples. The exponent differs from the value $0.784\pm0.003$ found in \cite{2009A&A...506..465H}. The apparent discrepancy can be explained by the fact that we consider here a significantly larger data set with lower error bars and that we do not scale the relation to the solar values of $\numax$ and $\dnumoy$. We note that the exponent also differs from \cite{2009MNRAS.400L..80S}, who found  $0.772\pm 0.005$.
This difference is not surprising since their fit is not based on red giants only but also includes main-sequence stars. Since the physical law explaining the relation between $\dnumoy$ and $\numax$ is not fully understood, one cannot exclude that the different properties of the interior structure between giants and dwarfs may explain the difference; we discuss this point in more detail below (Sect.~\ref{masserayon}). On the other hand, the independent fits based on the data in LRc01 and LRa01 analyzed separately give convergent results, with exponent respectively equal to 0.745 and 0.751.

The $\nde$ stars presented in Fig.~\ref{numax_dnu} were selected with a $\ampmax$ factor greater than the threshold level 8 defined in \cite{2009A&A...508..877M}. We verified that the 1-$\sigma$ spread of the data around the fit given by Eq.~(\ref{eqt_scaling}) is low, about 9\,\%. Thanks to the isomass lines superimposed on the plot, derived from the estimates  presented in Sect.~\ref{masserayon}, we remark that the spread in the observed relation between $\dnumoy$ and $\numax$ is mainly due to stellar mass. The metallicity dependence may also contributes to the spread; examining this effect is beyond the scope of this paper.

We have examined the few cases that differ from the fit by more than 20\,\%. As indicated in \cite{2009A&A...503L..21M}, they could correspond to a few halo stars (with large separation slightly above the main ridge of Fig.~\ref{numax_dnu}) or to higher mass stars (with large separation below the main ridge). We are confident that the possible targets with misidentified parameters in Fig.~\ref{numax_dnu} do not significatively influence the distributions and the fit. The analysis presented below, that gives the seismic measure of the stellar mass and radius, allows us to exclude outliers showing unrealistic stellar parameters: less than 2\,\%.

Histograms of the distribution of the seismic parameters $\numax$ and $\dnumoy$ have been plotted in Fig.~\ref{histo_numax} and Fig.~\ref{histo_dnu}. Deficits in the $\numax$ distribution around the diurnal frequencies of 11.6 and 23.2\muHz\ are related to corrections motivated by the spurious excess power introduced by the CoRoT orbit. As these artifacts have no fixed signature in $\dnumoy$, they are spread out, hence not perceptible, in the $\dnumoy$ distribution. The red-clump signature is easily identified as the narrow peak in the distribution of the mean large separation, around 4\muHz. The peak of the distribution of the maximum amplitude frequency is broader, with a maximum at 30\muHz\ and a shoulder around 40\muHz. This is in agreement with the synthetic population distribution \citep{1999MNRAS.308..818G, 2009A&A...503L..21M}.

\begin{figure}
\centering
\includegraphics[width=8.5cm]{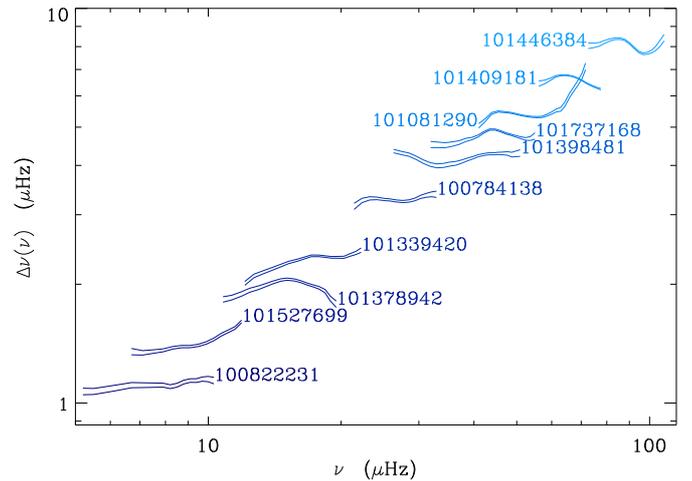}
\caption{$\dnu(\nu)$ relation for a set of red giants with a mass derived from the seismic parameters in the range [1.3, 1.4\,$M_\odot$]. Each pair of curves corresponds to a given red giant and gives the 1-$\sigma$ error bar. The numbers correspond to the identifications in the CoRoT data base.
\label{geante_corot_varidnu}}
\end{figure}

\subsection{Variation of the large separation with frequency\label{varidnu}}

The EACF allows us to examine the variation with frequency of the large separation ($\deltanunu$) and to derive more information about the stellar interior structure than given by the mean value. Significant variation of $\deltanunu$ is known to occur in the presence of rapid variation of the density, the  sound-speed and/or the adiabatic exponent $\Gamma_1$.

We have selected targets with similar mass, in the range [1.3, 1.4\,$M_\odot$], as inferred from the relation discussed in Sect.~\ref{masserayon}, but for increasing values of $\numax$. The corresponding $\deltanunu$ as a function of $\numax$ is plotted in Fig.~\ref{geante_corot_varidnu}. This allows us to examine how the global seismic signature evolves with stellar evolution. We remark that the large separation $\deltanunu$ exhibits a significant modulation or gradient for nearly all of these stars.

We note here that the variation of the large separation with frequency increases the uncertainty in the determination of $\dnumoy$ and the dispersion of the results. However, except for a few stars where the asymptotic pattern seems highly perturbed, we have confirmed that the measurement of $\dnumoy$ gives a good indication of the mean value of $\deltanunu$ over the frequency range where excess power is detected. The statistical analysis of $\deltanunu$ is beyond the scope of this paper and will be carried out in a further work.

\citet{2009A&A...508..877M} have shown that the analysis at high frequency resolution of $\deltanunu$ makes possible the identification of the mode degree in main-sequence stars observed with a high enough signal-to-noise ratio. This however seems ineffective for red giants, due to the fact that the oscillation pattern observed in red giants \citep{2010A&A...509A..73C, Barban2010} differs from the pattern observed for sub-giant and dwarf stars.



\begin{figure}
\centering
\includegraphics[width=8.5cm]{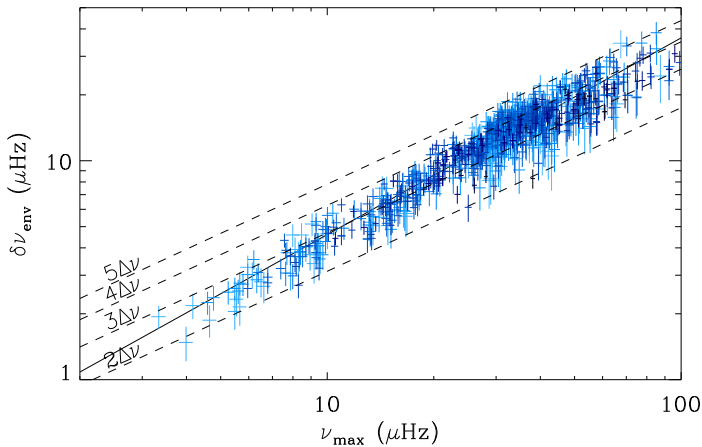}
\caption{$\numax$ - $\nuenv$ relation, with the same colour code as Fig.~\ref{numax_dnu}.
The solid line corresponds to the fit. The correspondence in units of the large separation is given by the dashed lines, which are guidelines for translating the excess power envelope width in large separation units.
\label{numax_env}}
\end{figure}

\begin{figure}
\centering
\includegraphics[width=8.5cm]{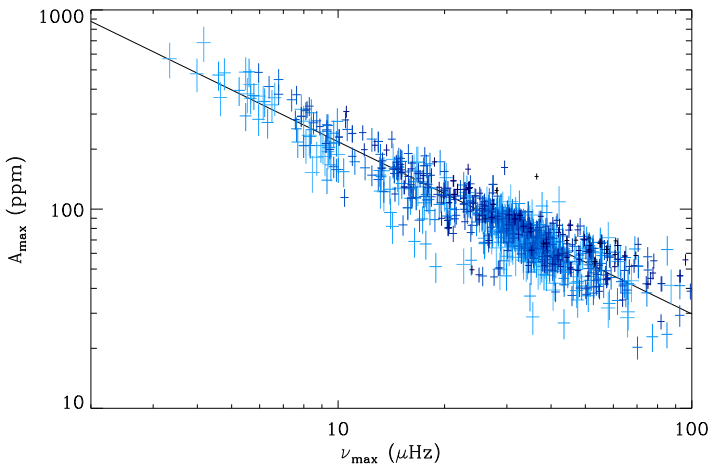}
\caption{$\numax$ - $\Abol$ relation. Same colour code as Fig.~\ref{numax_dnu}. The solid line indicates the best fit.
\label{numax_Abol}}
\end{figure}

\begin{figure}
\centering
\includegraphics[width=8.5cm]{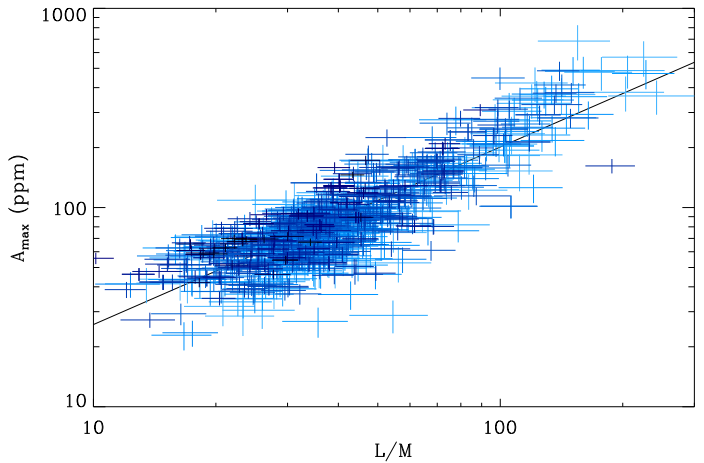}
\caption{$\Abol$ - $L/M$ relation. The fit assumes all $\Teff$ to be equal to the mean value.
Same colour code as in Fig.~\ref{numax_dnu}.
\label{geante_abol_lum_masse}}
\end{figure}

\section{Oscillation excess power\label{power}}

In this Section, we analyse the statistical properties of the parameters defining the excess power. They were measured for $\ntr$ targets with the highest signal-to-noise ratio, the excess power envelope being derived from a smoothed power spectrum.

\subsection{Excess power envelope\label{envelope}}

The full-width at half-maximum of the excess power envelope, plotted as a function of $\numax$ in Fig.~\ref{numax_env}, can be related to $\numax$:
\begin{equation}\label{eqt_env}
   \nuenv \simeq (0.59 \pm 0.02)\ \numax^{0.90\pm0.01}
\end{equation}
with $\numax$ in \muHz. As in Eq.~(\ref{eqt_scaling}) and for the same reason, we only report conservative error bars, about two times the error bars on $\numax$.
Combining Eq.~(\ref{eqt_scaling}) and (\ref{eqt_env}), we can derive the ratio between the envelope width and the mean large separation:
\begin{equation}\label{eqtenv}
   {\nuenv \over \dnumoy} \simeq (2.08 \pm 0.01)\ \numax^{0.15\pm0.02}
\end{equation}
with $\numax$ in \muHz.
This ratio is closely related to the number of observable peaks. Due to the small exponent, it does not vary significantly with $\numax$. However, we remark that the envelope width is narrower than 3 $\dnumoy$ when $\numax < 10$\muHz. Extending the validity of this relation up to solar-like stars does not seem possible, since the envelope width represents approximately $3.5\,\dnumoy$ for a red giant in the red clump but $10\,\dnumoy$ for a main-sequence star \citep{2009A&A...508..877M}. A major modification of the multiplicative parameter and/or of the exponent of the scaling law may occur as the stellar class changes.

Measurements at frequency above 100\muHz\ will help to improve the exact relation $\nuenv (\numax)$. We have noted that the measurement of the envelope width is sensitive to the method, to the low-pass filter applied to the spectrum, if any, and to the estimate of the background.
Since smoothing or averaging with a large filter width is inadequate for red giants with narrow excess power envelopes, and since an inadequate estimate of the background translates immediately to a biased determination of $\nuenv$, we used a narrow smoothing, of the order of $1.5\;\dnumoy$.

We have also directly estimated the number of eigenmodes with an H0 test \citep{2004A&A...428.1039A}. Data were rebinned over 5 pixels. Selected peaks were empirically identified to the same mode if their separation in frequency is less than $\dnumoy / 30$. The median number $\npeak$ of peaks detected as a function of frequency is given in Table~\ref{npic}. We note that this number does not vary along the spectrum, in agreement  with the small exponent of Eq.~(\ref{eqtenv}).

Estimates of the minimum and maximum orders of the detected peaks were simply obtained by dividing the minimum and maximum eigenfrequencies selected with the H0 test by $\dnumoy$. They vary in agreement with the exponent given by Eq.~(\ref{eqtenv})
(Table~\ref{npic}).

\begin{table}
\caption{Number of detected peaks}\label{npic}
\begin{tabular}{ccccc}
\hline
\multicolumn{2}{c}{$\nu$ range} & $\bar{n}\ind{min}$ & $\bar{n}\ind{max}$ & $\npeak$ \\
\multicolumn{2}{c}{(\muHz)} & \\
\hline
&$\le$15  & 4.1 & 9.8 &  8\\
 15 & 20  & 5.0 &11.0 &  8\\
 20 & 30  & 5.5 &11.3 &  7\\
 30 & 40  & 6.1 &11.9 &  7\\
 40 & 60  & 7.0 &13.1 &  7\\
 60 &100  & 8.3 &15.1 &  7\\
100 &300  & 8.9 &15.9 &  8\\
\hline
\end{tabular}

For each frequency interval, $\bar{n}\ind{min}$ and $\bar{n}\ind{max}$ are the median values of the minimum and maximum detected eigenfrequencies expressed in unit $\dnumoy$; $\npeak$ is the median of the number of detected peaks, at a rejection level of 10\,\%.
\end{table}

\subsection{Temperature\label{temperatures}}

Information on effective temperature is required to relate the oscillation amplitude to interior structure parameters. The effective temperatures have been derived from dereddened 2MASS colour indices using the calibrations of \cite{1999A&AS..140..261A}, as described in \cite{Baudin2010} for stars in LRc01. For the 3 stars without 2MASS data, optical $BVr^{\prime}i^{\prime}$ magnitudes taken from Exo-Dat have been used \citep{2009AJ....138..649D}. We have adopted $A_V$ = 0.6 mag for LRa01 based on the extinction maps of \cite{2005PASJ...57..417D} and \cite{2009MNRAS.395.1640R}. As for LRc01, the good agreement between the $\Teff$ values derived from near-IR and optical data indicates that this estimate is appropriate. The statistical uncertainty on these temperatures is of the order of 150\,K considering the internal errors of the calibrations and typical uncertainties on the photometric data, reddening and metallicity. Turning to the systematic uncertainties, employing other calibrations would have resulted in temperatures deviating by less than 150\,K \citep{1999A&AS..140..261A}.

A clear correlation between $\Teff$ and $\numax$ is apparent:
\begin{equation}\label{numax_temp}
T \propto \numax^{0.04\pm 0.01}.
\end{equation}

\subsection{Amplitude\label{bolamp}}

Maximum amplitudes of radial modes have been computed according to the method proposed by
\cite{2009A&A...495..979M} for CoRoT photometric measurement. The distribution of the maximum mode amplitude  as a function of $\numax$, presented in Fig.~(\ref{numax_Abol}), is:
\begin{equation}\label{numax_abol}
   \Abol  \simeq (1550 \pm 100)\ \numax^{-0.85 \pm 0.02}
\end{equation}
with $\Abol$ in parts-per-million (ppm) and again with  $\numax$ in \muHz. The median relative dispersion is of about 50\,\%. Again, the fits based on LRc01 and LRa01 data separately are equivalent. In order to avoid biasing the exponent with data presenting a gradient of signal-to-noise ratio with frequency, we have estimated the exponents for subsets of stars with similar signal-to-noise ratios. This proves to be efficient since we then get convergent results for the exponent in Eq.~(\ref{numax_abol}).

Using several 3D simulations of the surface of main-sequence stars, \cite{2007A&A...463..297S} have found that the maximum of the mode amplitude in \emph{velocity} scales as $(L/M)^{s}$ with $s=0.7$.  This scaling law reproduces rather well the  main-sequence stars observed so far in Doppler velocity from the ground. When extrapolated to the red-giant domain ($L/M \ge 10$), this scaling law results in a good agreement with the  giant and sub-giant stars observed in
Doppler velocity.
To derive the mode amplitude in terms of bolometric intensity fluctuations from the mode amplitude in velocity, one usually assumes the adiabatic relation proposed by \cite{1995A&A...293...87K}. For the mode amplitudes in intensity, this gives a scaling law of the form  $(L/M)^s\, \Teff^{-1/2}$, which requires the measurement of effective temperatures. Due to the Stefan-Boltzmann law, $L/M$ scales as $\Teff^4 / g$, hence as $\Teff^{7/2} / \numax$.

As a consequence of Eq.~(\ref{numax_temp}), $\Teff^{7/2} / \numax$ does not scale exactly as $\numax^{-1}$. We then obtain the scaling of the amplitude with $(L/M)^s\, T^{-1/2}$ (Fig.~\ref{geante_abol_lum_masse}):
\begin{equation}\label{eqt_abol}
  \Abol \propto \ \left({L\over M}\right)^{0.89\pm0.02} \Teff^{-1/2}.
\end{equation}
The spread around the global fit is as large as for the relation $\Abol (\numax)$ (Eq.~\ref{numax_Abol}).
The influence of  $\Teff$  in Eq.~(\ref{eqt_abol}) gives an exponent $s$ that is different from the opposite of the exponent in Eq.~(\ref{numax_abol}) as would be the case if all temperatures were fixed to a single mean value. Finally, we derive an exponent of the scaling law between the maximum amplitude and the ratio $L/M$, 0.89$\pm$0.02, which differs significantly from the 0.7 value found for main-sequence stars and subgiants observed in velocity \citep{2007A&A...463..297S}.

\begin{figure}
\centering
\includegraphics[width=8.5cm]{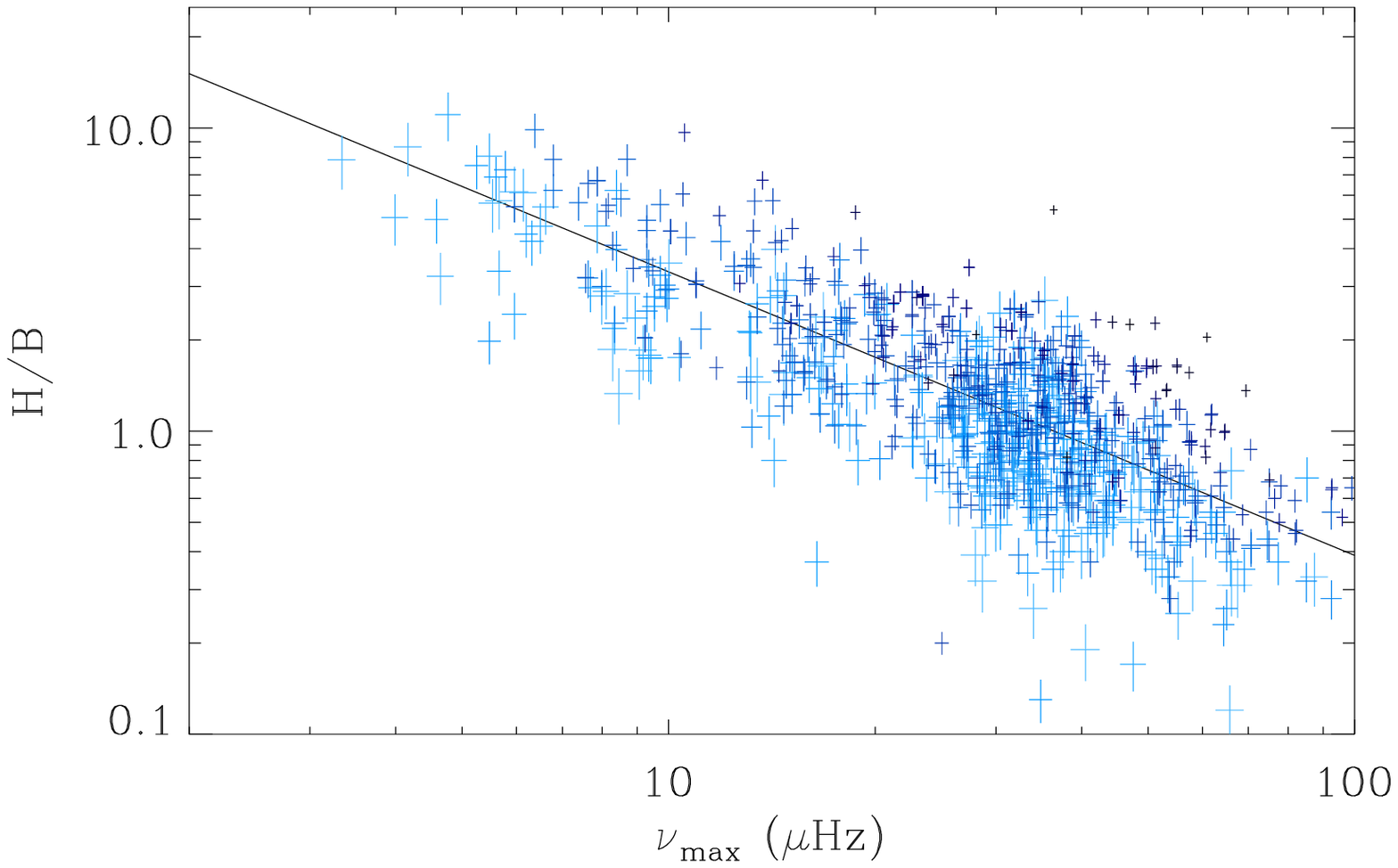}
\caption{$\numax$ - $\HBR$ relation. Same colour code as in Fig.~\ref{numax_dnu}.
\label{numax_HBR}}
\end{figure}

\subsection{Height-to-background ratio\label{height}}

Due to the large variety of stellar activity within the red-giant data set, the ratio $\HBR$ does not obey a tight relation, but increases when $\numax$ decreases (Fig.~\ref{numax_HBR}). With the same method as for the amplitude, we derive the relation $\HBR \propto \numax^{-0.98 \pm 0.05}$. This indicates first that it is possible to measure oscillation with a large height-to-background ratio at very low frequency, which is encouraging for future very long observations as will be provided by Kepler.

The comparison of the mode amplitude and of the height-to-background ratio with frequency shows that the mean amplitude of granulation and activity scales as $\numax^{-0.36}$. This can be compared to Eq.~(\ref{numax_abol}) with an exponent of about $-0.85$. If we link the amplitude to the fraction of the convective energy injected in the oscillation, we conclude that this fraction is greater at low frequency. Furthermore, even if less convective energy is injected in the oscillation compared to granulation, the fraction injected in the oscillation increases more rapidly at low frequency than the fraction injected into the granulation.

\begin{figure}
\centering
\includegraphics[width=8.5cm]{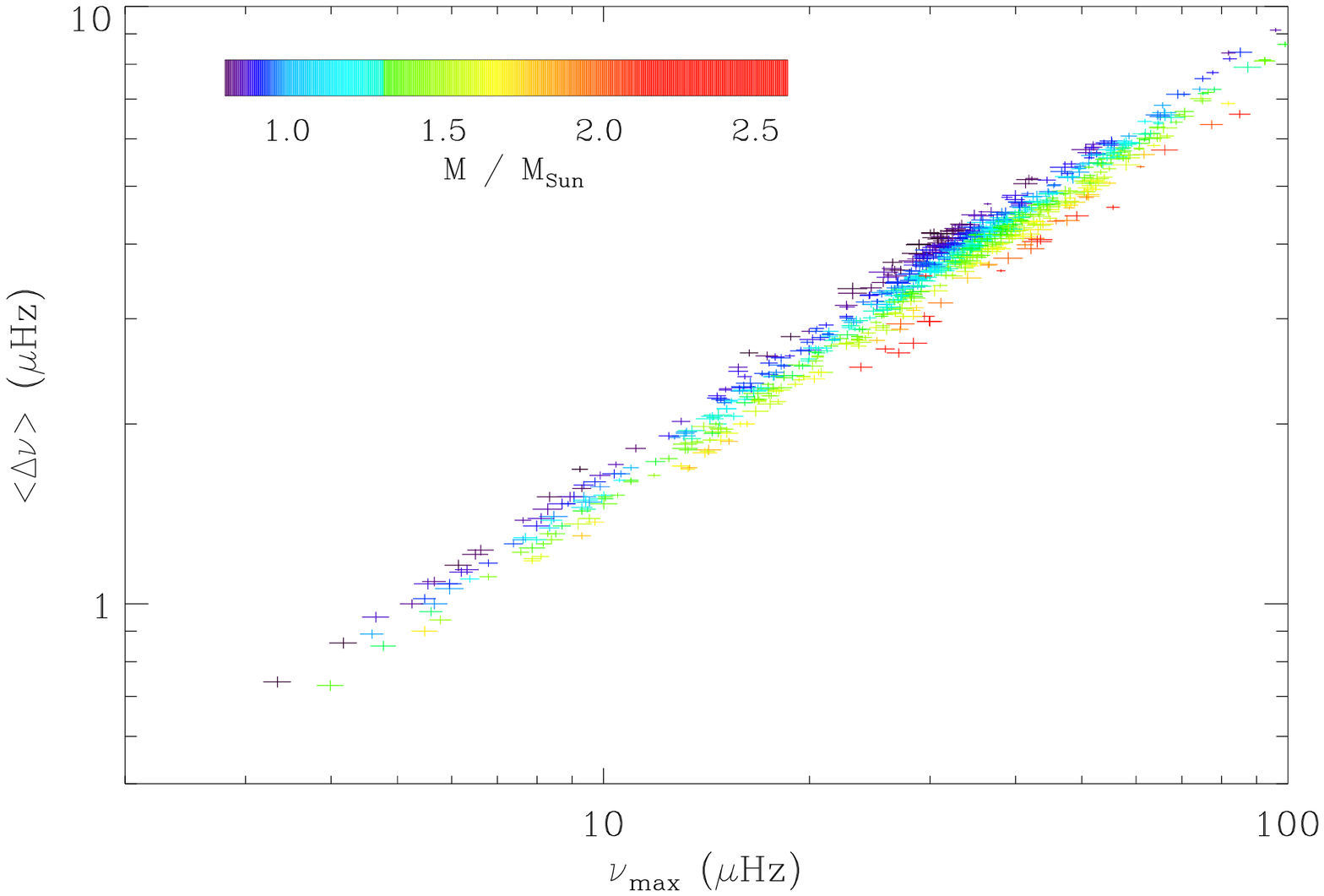}
\caption{$\numax$ - $\dnumoy$ relation, restricted to the $\ntr$ targets.
The colour code relates the stellar mass derived from Eq.~(\ref{masse}).
\label{numax_dnu_masse}}
\end{figure}

\begin{figure}
\centering
\includegraphics[width=8.5cm]{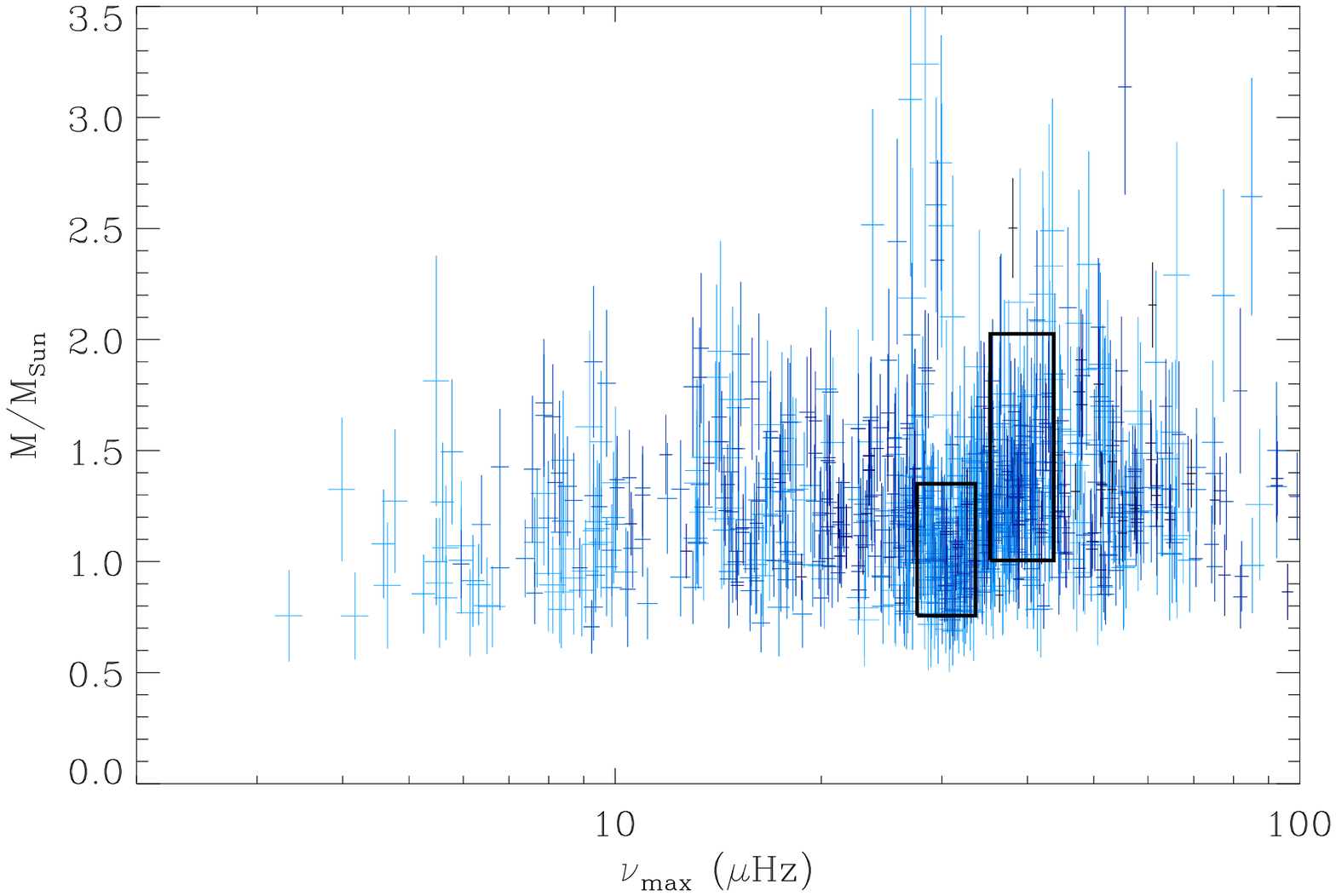}
\includegraphics[width=8.5cm]{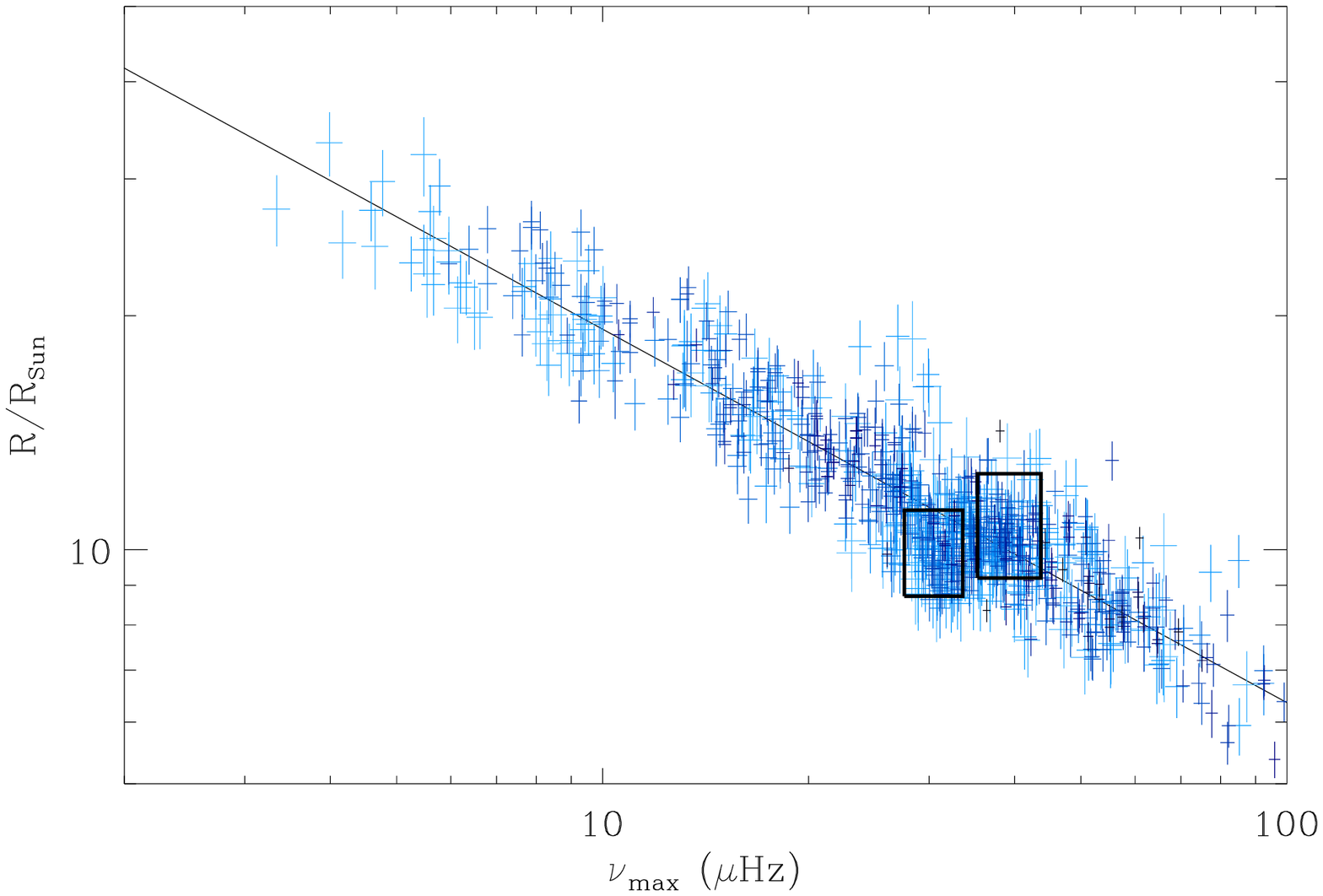}
\caption{$\numax$ - $M$ and $\numax$ - $R$ relations. Same colour code as in Fig.~\ref{numax_dnu}. The black rectangles delimit the two components of the red clump identified in Figs.~\ref{histo_numax} and \ref{compare_run_histo}.
\label{numax_MR}}
\end{figure}

\begin{figure}
\centering
\includegraphics[width=8.5cm]{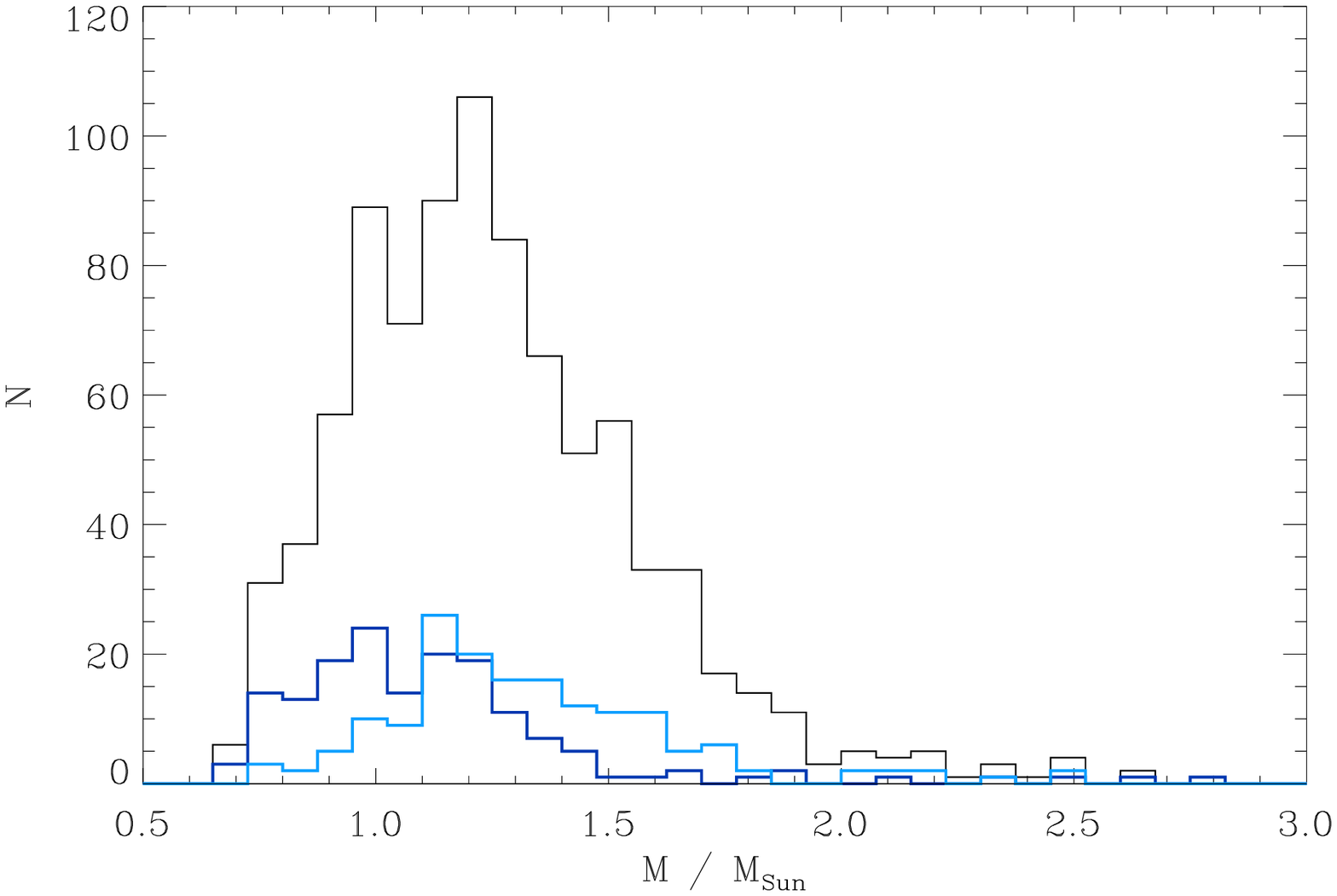}
\includegraphics[width=8.5cm]{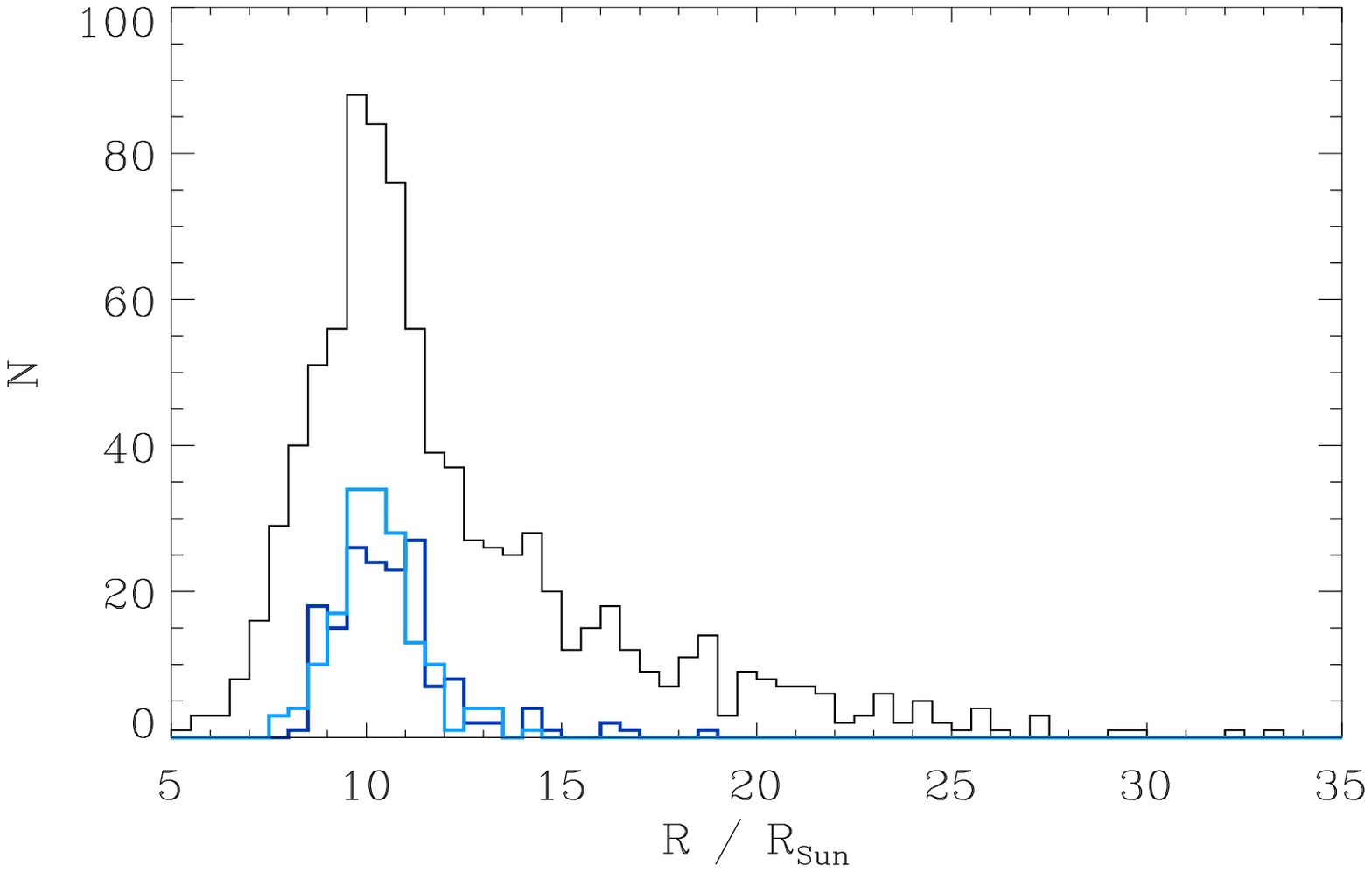}
\caption{Histograms of the stellar mass and radius. The curves in dark and light blue correspond respectively to the components of the red clump around 30 and 40\muHz. The main difference appears on the mass distribution: 35\,\% higher for the component at 40\muHz.
1-$\sigma$ uncertainties at the red clump location are typically $0.24\,M_\odot$ and $0.8\,R_\odot$.
\label{histo_MR}}
\end{figure}

\begin{table*}
\caption{Calibration of the red-giant mass and radius}\label{scaling_MR}
\begin{tabular}{lrrrrrrrl}
\hline
star & $\Teff$  & $\numax$ & $\dnumoy$ & $R\ind{mod}$ & $\Rsis$& $M\ind{mod}$ & $\Msis$ & Ref.\\
     & (K)      & \multicolumn{2}{c}{\dotfill (\muHz)\dotfill}& \multicolumn{2}{c}{\dotfill $R_\odot$\dotfill}& \multicolumn{2}{c}{\dotfill $M_\odot$\dotfill}\\
\hline
HD181907         & 4760$\pm$120&  29.1$\pm$0.6& 3.47$\pm$0.1&  12.3$\pm$0.6 &  12.3& 1.2 $\pm$0.3& 1.53&  (1)  \\
$\beta$       Oph& 4550$\pm$100&  46.0$\pm$2.5&  4.1$\pm$0.2& 12.2 $\pm$0.8 &  13.8& 2.90$\pm$0.3& 2.99&  (2)  \\
$\varepsilon$ Oph& 4850$\pm$100&  53.5$\pm$2.0&  5.2$\pm$0.1& 10.55$\pm$0.15&  10.2& 1.85$\pm$0.05& 1.91&  (3) \\
$\xi$   Hya      &4900$\pm$75\ &  92.3$\pm$3.0&  7.0$\pm$0.2& 10.4 $\pm$0.5 &  9.70& 3.04$\pm$0.1& 3.01&  (4)    \\
$\eta$  Ser      & 4900$\pm$100& 125.0$\pm$3.0& 10.1$\pm$0.3&  6.3 $\pm$0.4 &  6.31& 1.60$\pm$0.1& 1.73&  (5)    \\
\hline
\end{tabular}

The columns $R\ind{mod}$ and $M\ind{mod}$ indicate the radius and mass derived from a detailed models, whereas $\Rsis$ and $\Msis$ indicate the radius and mass derived from Eqs. \ref{rayon} and \ref{masse}.


(1) \cite{2010A&A...509A..73C}, \cite{Miglio2010}

(2) \cite{2009ASPC..404..307K}

(3) \cite{2008A&A...478..497K}, \cite{2009A&A...503..521M}, \cite{2007A&A...468.1033B},

(4) \cite{2002A&A...394L...5F}

(5) \cite{2004ESASP.559..113B},  \cite{2006A&A...458..931H}, and F. Carrier, private communication

More references on the targets are given in \cite{2009ASPC..404..307K}
\end{table*}

\begin{table*}
\caption{Scaling with $\numax$}\label{nuc_numax}
\begin{tabular}{lllll}
\hline
$f(\numax)$       & $\dnumoy$            & $\Teff$         & $\Rsis$       &  $\nuc$        \\
exponent          & $\beta$              & $\gamma$        & $\delta$      &  $\alpha$      \\
\hline
    giants        & 0.75$\pm$0.01        &\ \ 0.04$\pm$0.01& $-$0.48$\pm$0.01& 1.00$\pm$0.02 \\
sub-giants$^{(1)}$& 0.71$\pm$0.02        & $-$0.02$\pm$0.02& $-$0.58$\pm$0.06& 0.83$\pm$0.10 \\
                  & 0.77$\pm$0.01$^\star$&                 &               &   0.97$\pm$0.10 \\
dwarfs$^{(2)}$    & 0.76$\pm$0.02        & $-$0.18$\pm$0.02& $-$0.64$\pm$0.03& 0.97$\pm$0.04 \\
                  & 0.77$\pm$0.01$^\star$&                 &               &   0.99$\pm$0.04 \\
\hline
\end{tabular}

References:

(1) \cite{2009ASPC..404..307K}, \cite{Deheuvels2010}

(2) \cite{2009ASPC..404..307K}, \cite{2008A&A...488..635M}, \cite{2009A&A...494..237T},  \cite{Ballot2010},  \cite{Mathur2010}, \cite{Gaulme2010}

$\star$ value from \cite{2009MNRAS.400L..80S}

\end{table*}
\section{Red-giant mass and radius estimate\label{masserayon}}

It is possible to derive the stellar mass and radius from $\dnumoy$ and $\numax$, as done for example by \cite{2009ASPC..404..307K} for a few giant targets.
Such a scaling supposes that the mean large separation $\dnumoy$ is proportional to the mean stellar density and that $\numax$ varies linearly with the cutoff frequency $\nuc$, hence with $g/\sqrt{\Teff}$, $g$ being the surface gravity
\citep{1991ApJ...368..599B}. The analysis is extended here to a much larger set of targets, and measurements are improved compared to \cite{2009ASPC..404..307K} since we have introduced the individual stellar effective temperatures.

Before any measurements, we have calibrated the scaling relations giving the stellar mass and radius as a function of the asteroseismic parameters by comparing the seismic and modeled mass and radius of red giants with already observed solar-like oscillations (Table \ref{scaling_MR}):
\begin{eqnarray}
  {\Rsis \over\Rs}  &=& r\ \ \,\left({\numax \over \numaxs}\right) \
     \left({\dnu \over \dnus}\right)^{-2}
     \left({T \over \Ts}\right)^{1/2} \label{rayon}\\
  {\Msis\over\Ms} &=& m\ \left({\numax \over \numaxs}\right)^{3}
     \left({\dnu \over \dnus}\right)^{-4} \left({T \over \Ts}\right)^{3/2} \label{masse}
\end{eqnarray}
According to the targets summarized in Table \ref{scaling_MR}, the factors $r$ and $m$ are 0.90$\pm$0.03 and 0.89$\pm$0.07 respectively, for $\dnus =135.5$\muHz, $\numaxs = 3050$\muHz\ and $\Ts=5\,777$\,K.
When taking into account these factors, the agreement between the modeled and seismic values of the radius and mass of the targets listed in Table~\ref{scaling_MR} is better than 7\,\% and agrees within the error bar of the modeling.
Calculations were only performed for the $\ntr$ targets with the best signal-to-noise ratios (Fig.~\ref{numax_dnu_masse}).
The error bars in $R$ and $M$ inferred for the CoRoT red giants from Eqs.~(\ref{rayon}) and (\ref{masse}) are about 8 and 20\,\% (Fig.~\ref{numax_MR}).

The equation giving the mass is highly degenerate, since $\numax^3 \dnu^{-4}$ is nearly constant according to Eq.~(\ref{eqt_scaling}). This degeneracy shows that the temperature strongly impacts the stellar mass. It also indicates that the dispersion around the scaling relation (Eq.~\ref{eqt_scaling}) is the signature of the mass dispersion.

We derive from  Fig.~(\ref{numax_MR}) the relation between the stellar radius and $\numax$:
\begin{equation}\label{numax_radius}
    {\Rsis\over R_\odot} = (56.7\pm 1.0) \ \numax^{-0.48\pm0.01}
\end{equation}
As in previous similar equations, $\numax$ is expressed in \muHz\ and error bars are conservative.


We have analyzed this result to examine to what extent the exponent close to $-1/2$ is due the dependence of the cutoff frequency to the gravity field $g$ and to explicit the relation between $\numax$ and $\nuc$. In order to perform this in detail and to understand the difference reported in Eq.~(\ref{eqt_scaling}) compared to \cite{2009MNRAS.400L..80S}, we  assume a variation of the cutoff frequency with $\numax$ as $\nuc \propto \numax^\alpha$ and then reapply Eq.~(\ref{rayon}) and (\ref{masse}), taking into account the scalings $\dnumoy\propto \numax^\beta$ and $\Teff \propto \numax^\gamma$. We obtain a new relation
$\Rsis \propto \numax^\delta$:
\begin{equation}\label{rayon_bis}
    \Rsis \propto \numax^{\alpha - 2\beta + \gamma/2}
\end{equation}
which has to coincide with (Eq.~\ref{numax_radius}). Then, when we introduce the numerical values found for the exponents of the different fits, the comparison of Eq.~(\ref{rayon_bis}) with Eq.~(\ref{numax_radius}) gives an exponent $\alpha$ very close to 1, within 2\,\%. This demonstrates that the ratio $\numax / \nuc$ is constant, which was widely assumed but is verified for the first time for red giants. Its value is about 0.64.

\begin{figure}
\centering
\includegraphics[width=8.5cm]{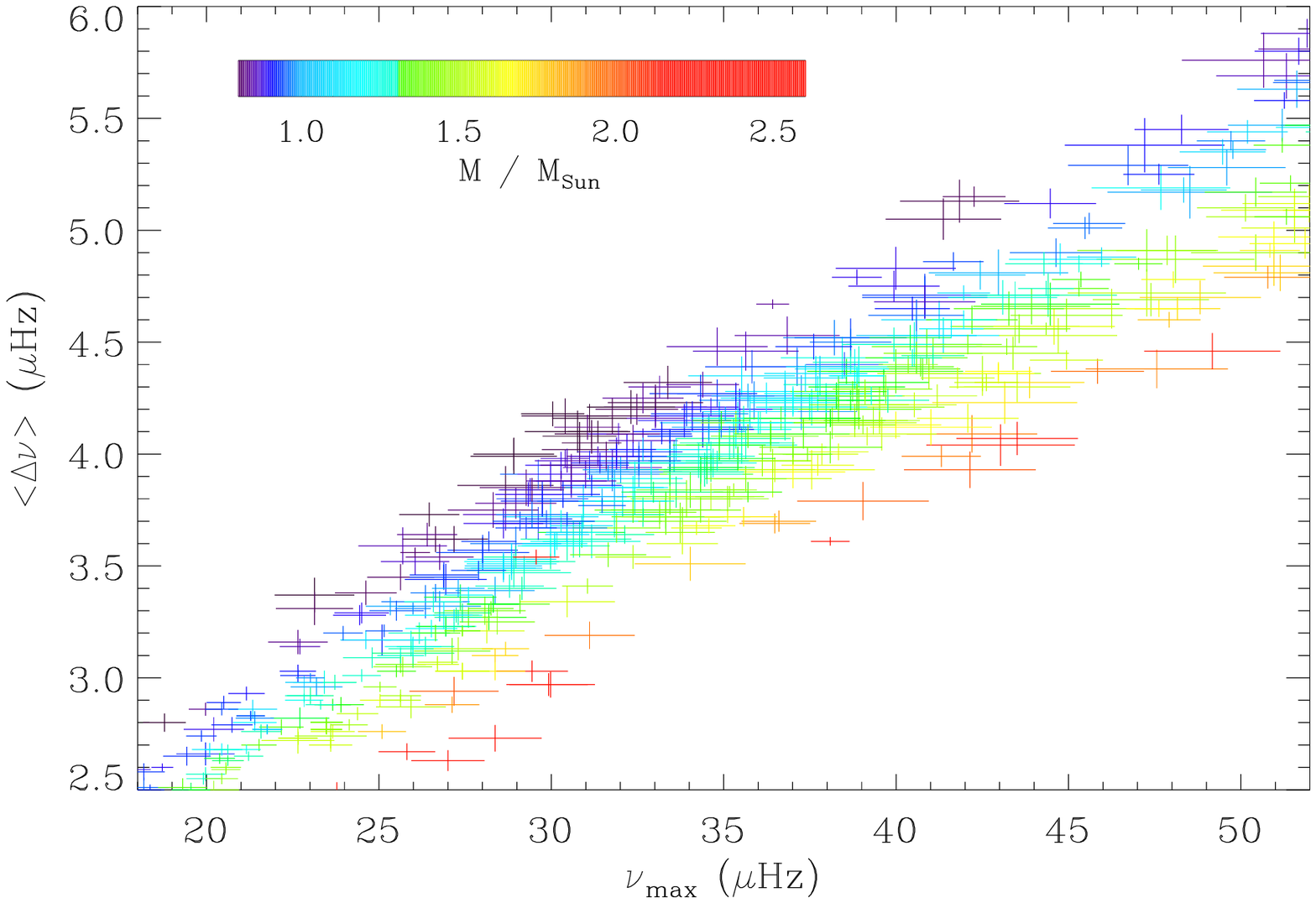}
\caption{$\numax$ - $\dnumoy$ relation around the red clump. Same colour code for the mass as in  Fig.~\ref{numax_dnu_masse}. Low-mass giants are located in the main component of the red clump around 30\muHz.
\label{numax_dnu_clump}}
\end{figure}

From Table 1 of \cite{2009ASPC..404..307K} completed with a few other solar-targets benefitting from a precise modeling (HD 203608, \cite{2008A&A...488..635M}; $\iota$ Hor, \cite{2009A&A...494..237T}; HD 52265, \cite{Ballot2010}; HD 170987, \cite{Mathur2010}; HD 46375, \cite{Gaulme2010}), we can derive the exponent $\alpha$ from the values of  $\beta$, $\gamma$ and $\delta$ for main-sequence stars and for sub-giants. In spite of the quite different set of exponents, we infer in both cases $\alpha\simeq 1$, namely a relation between $\nuc$ and $\numax$ very close to linear (Table~\ref{nuc_numax}).
This proves that for all stellar classes the assumption of a fixed ratio $\numax / \nuc$ is correct. Error bars for sub-giants and main-sequence stars are larger compared to the red-giant case due to the limited set of stars, and maybe also due to inhomogeneous modeling.


\begin{table}
\caption{Distribution of the red-clump parameters}\label{MRclump}
\begin{tabular}{llcccc}
\hline
Freq. &$\#$ of& $\Rsis/ R_\odot$    & $\Msis / M_\odot$      & $L / L_\odot$          & $\Teff$               \\
(\muHz)& stars&                     &                        &                        & (K)                   \\
\hline
28-32 & 146   & 10.0$_{-1.3}^{+1.1}$& 1.00$_{-0.18}^{+0.19}$ & 38.5$_{-10.0}^{+9.7} $ & 4510$_{-180}^{+170} $ \\
      & LRc01  & 10.1$_{-0.7}^{+0.7}$& 1.01$_{-0.15}^{+0.17}$ & 39.0$_{-7.0 }^{+9.5} $ & 4530$_{-150}^{+140} $ \\
      & LRa01  &  9.8$_{-0.6}^{+0.7}$& 0.98$_{-0.12}^{+0.20}$ & 38.1$_{-7.0 }^{+8.5} $ & 4480$_{-100}^{+120} $ \\
\hline
37-43 & 137   & 10.3$_{-0.7}^{+1.1}$& 1.34$_{-0.18}^{+0.25}$ & 42.4$_{- 8.3}^{+11.2}$ & 4580$_{-170}^{+140} $ \\
      & LRc01  & 10.3$_{-0.7}^{+0.8}$& 1.33$_{-0.15}^{+0.22}$ & 42.1$_{- 7.8}^{+9.4}$  & 4580$_{-140}^{+130} $ \\
      & LRa01  & 10.4$_{-0.6}^{+1.2}$& 1.36$_{-0.19}^{+0.28}$ & 42.7$_{- 7.3}^{+8.3}$  & 4590$_{-150}^{+120} $ \\
\hline

\end{tabular}

The 2nd column gives first the total number of targets considered in each region of the clump, then indicates which field is considered.

The following columns present the main values of the stellar parameters and the dispersion of their distribution in each region
\end{table}

\section{Red-giant population\label{redclump}}

\subsection{Red clump}

\cite{2009A&A...503L..21M} have compared synthetic, composite stellar populations to CoRoT observations. The analysis of the distribution in $\numax$ and $\dnumoy$ allows them to identify red-clump giants and  to estimate the properties of poorly-constrained populations. Benefitting from the reduction of the error bars and from the extension of the analysis to lower frequencies compared to previous works, we can derive precise properties of the red clump (Fig.~\ref{numax_dnu_clump}).

The distribution of $\numax$ is centred around $30.2$\muHz\ with 69\,\% of the values within the range  $30.2\pm 2.0$\muHz.
The maximum of the distribution of $\numax$ is located at $29.7 \pm 0.2$\muHz.
The corresponding distribution of $\dnumoy$ is  centered around $3.96$\muHz\ with 69\,\% of the values within the range  $3.96\pm 0.33$\muHz, and its maximum located at $3.97 \pm 0.03$\muHz. A second contribution of the red clump can be identified around 40\muHz. This feature may correspond to the secondary clump of red-giant stars predicted by \cite{1999MNRAS.308..818G}.

Table~\ref{MRclump} presents the mean values and the distribution of the physical parameters identified for the peak and the shoulder of the clump stars: the mean values of the radius are comparable for the two components, but effective temperature, mass and luminosity are slightly different. The members of the second component are hotter by about 80\,K, brighter, and significantly more massive. The mass distribution disagrees with the theoretical prediction. The distribution is centered on 1.32\,$M_\odot$ (Fig.~\ref{histo_MR}) whereas \cite{1999MNRAS.308..818G} predicts 2-2.5\,$M_\odot$ for solar metallicity. Similarly, stars appear to be brighter, contrary to the theoretical expectations.
Figure~\ref{numax_dnu_clump} presents a zoom on the red-clump region of the $\dnumoy$ versus $\numax$ relation and shows that stars less massive than $1\,M_\odot$ are numerous in the main component of the clump but rare in the shoulder. Contrary to \cite{1999MNRAS.308..818G}, we do not identify many stars with a mass above 2\,$M_\odot$ near the second component of the clump.
Selecting stars in the secondary component of the clump by only a criterion on $\numax$ is certainly not sufficient, since many stars with $\numax$ around 40\,\muHz\ can belong to the tail of the distribution of the main component. The discrepancy with the prediction of \cite{1999MNRAS.308..818G} may result from the way that we identify the stars and a refined identification will be necessary to better describe this secondary component.

\begin{figure*}
\centering
\includegraphics[width=14cm]{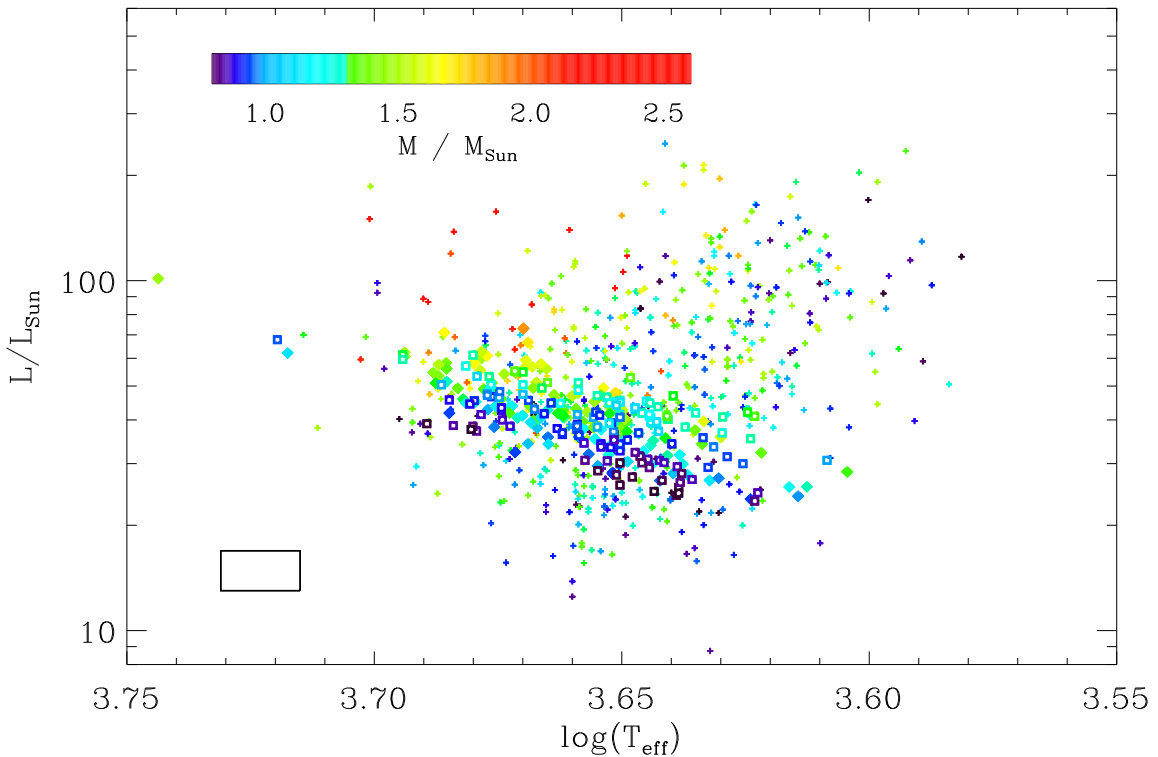}
\caption{HR diagram of the $\ntr$ targets of LRa01 and LRc01. The estimates of the mass are derived from Eq.~(\ref{masse}) and are presented with the same colour code as Fig.~\ref{numax_dnu_masse}.
For clarity, individual bars are not represented. The mean 1-$\sigma$ error box is given in the lower-left corner of the diagram. Cross are replaced by open squares for stars in the main component of the red clump, and diamonds for the second component.
\label{HR_clump}}
\end{figure*}

Figure~\ref{HR_clump} presents an HR diagram of the red-clump stars among all targets with precise asteroseismic parameters, with the stellar luminosity derived from the Stefan-Boltzmann law. We note a mass gradient in the direction of hot and luminous objects. However, we note that the members of the two components are intricately mixed in this diagram.

\begin{figure}
\centering
\includegraphics[width=8.5cm]{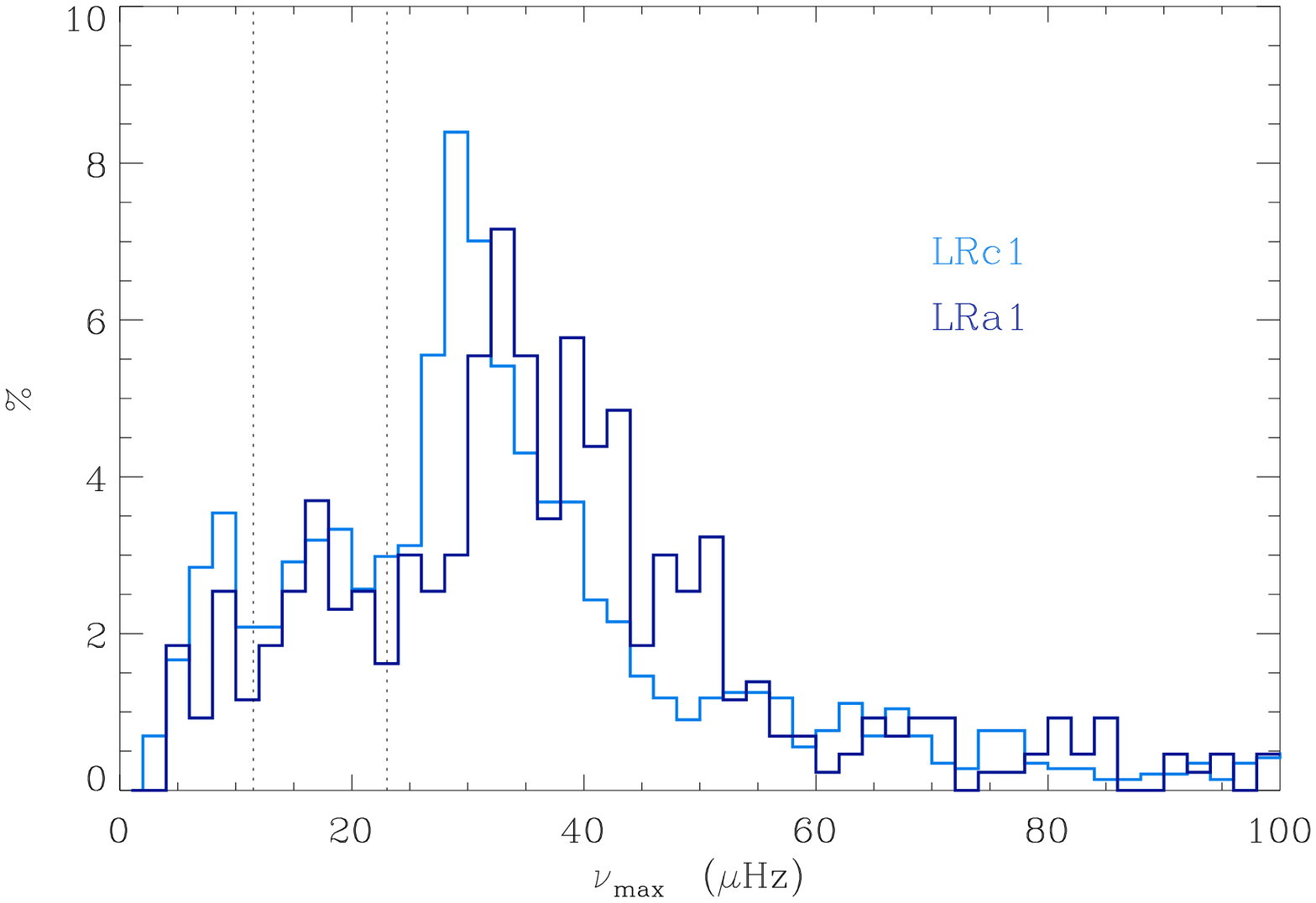}
\includegraphics[width=8.5cm]{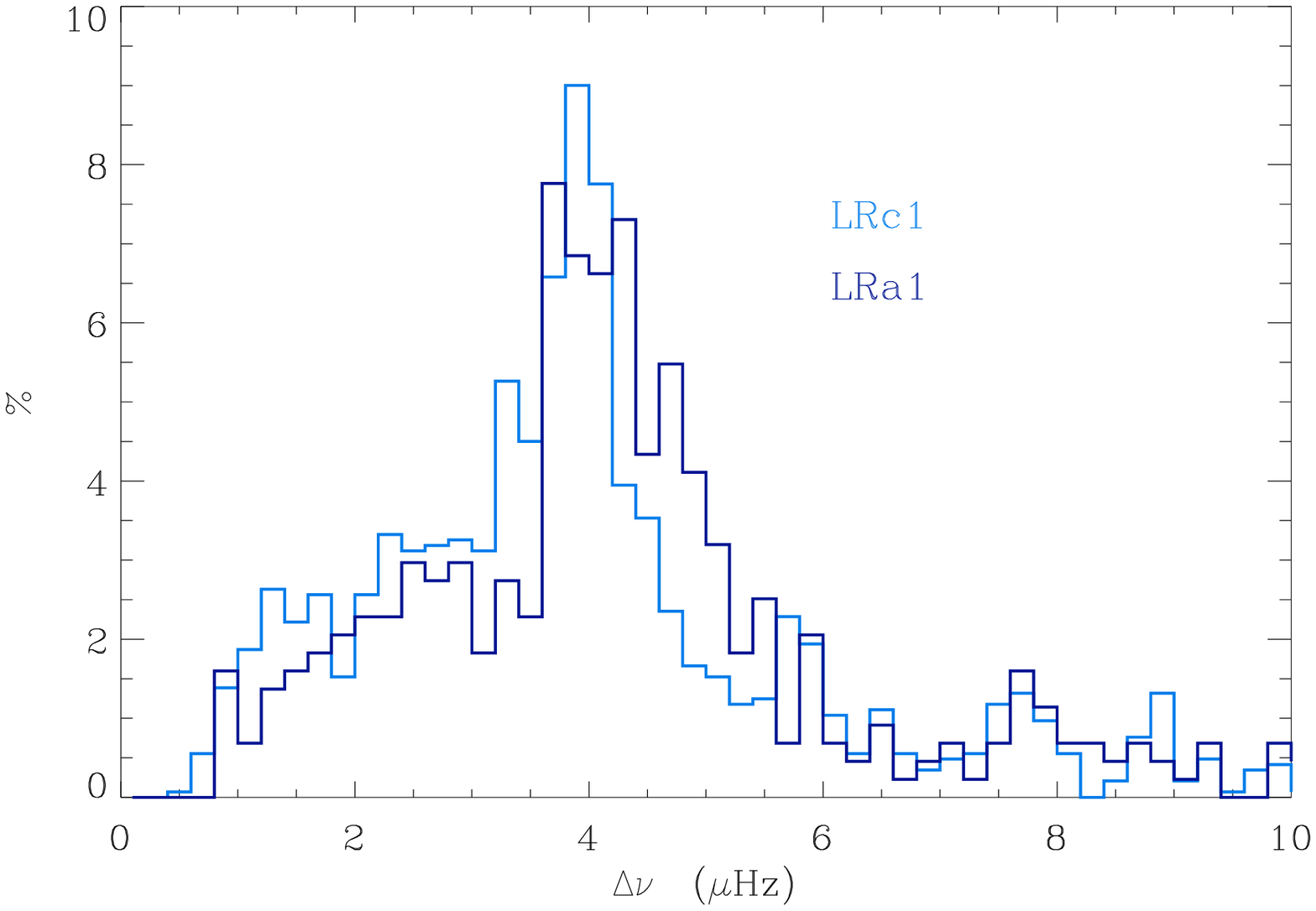}
\caption{Histograms of $\numax$ and $\dnumoy$, comparing the populations in LRc01 and LRa01.
The secondary red-clump signature is mainly due to the population in LRa01.
1-$\sigma$ uncertainties at the red clump are typically 0.06\muHz\ for the $\dnumoy$-axis, 1.0\muHz\ for the $\numax$-axis and 1\,\% on both y-axes.
As in Fig.~\ref{histo_numax}, the dotted lines in the histogram  of $\numax$indicate the deficits of reliable results around 11.6 and 23.2\muHz .
\label{compare_run_histo}}
\end{figure}

\subsection{Comparison center/anticenter}

We have compared the distribution of $\numax$ and $\dnumoy$ of the 2 CoRoT runs LRc01 and LRa01. The LRc01 run, centered on $(\alpha, \delta)$ = (19h25min, 0$^\circ$30\arcmin), points towards an inner region of the Galaxy: galactic longitude and latitude 37$^\circ$ and $-07^\circ$45\arcmin\ respectively, 38$^\circ$ away from the Galactic center.
Targets of the run LRa01 are located in the opposite direction of LRc01, centered on $(\alpha, \delta)$ = (6h42min, $-0^\circ$30\arcmin), with a lower galactic latitude ($-01^\circ$45\arcmin) and galactic longitude of 212$^\circ$. According to the reddening inferred for the targets, a typical 13th magnitude red giant of the red clump is located at 3\,kpc.

The histograms of $\numax$ and $\dnumoy$ for both fields are compared in Fig.~\ref{compare_run_histo}.
They show comparable relative values in all frequency ranges except the location of the clump stars. The main component of the red clump is much less pronounced in LRa01 compared to LRc01; on the other hand, the second component of the clump is more populated in LRa01. The distribution concerning LRa01, with two components, strongly supports the identification of the secondary clump.
Since the two populations were selected according to homogeneous criteria and show comparable scaling laws for all asteroseismic parameters, understanding the difference between them will require a further analysis taking into account more parameters than those given by asteroseismology, in order to investigate the role of evolutionary status, metallicity and position in the Galaxy.

\section{Conclusion\label{conclusion}}

We have shown the possibility of extracting statistical information from the high-precision photometric time series of a large sample of red giants observed with CoRoT and analyzed with an automated asteroseismic pipeline. We summarize here the main results as well as open issues:

- Out of more than 4\,600 time series, we have identified more than 1\,800 red giants showing solar-like oscillations. We have extracted a full set of precise asteroseismic parameters for more than 900 targets.

- Thanks to the detection method, we are able to observe precise large separations as small as 0.75\muHz.  We obtain reliable information for the seismic parameters  $\dnumoy$ and  $\numax$ for $\numax$ in the range [3.5, 100\muHz]. We have shown that the detection and measurement method does not introduce any bias for $\numax$ above 6\muHz. This allows us to study in detail the red clump in the range [30, 40\muHz].

- We have proposed scaling relations for the parameters defining the envelope where the asteroseismic power is observed in excess. We note that the relation defining the full-width at half-maximum $\nuenv$ of the envelope cannot be extended to solar-like stars. The scaling relation between $\nuenv$ and $\numax$ is definitely not linear for giants: $\nuenv \propto \numax^{0.90}$. The maximum amplitude scales as $\numax^{-0.85}$ or $(L/M)^{0.89}$. Deriving bolometric amplitudes will require more work: examination of the equipartition of energy between the modes and stellar atmosphere modeling.

- When supplemented by the effective temperature, the asteroseismic parameters $\dnumoy$ and $\numax$ give access to the stellar mass and radius. Red-giant masses derived from asteroseismology are degenerate, but it is still possible to estimate their value with a typical uncertainty of about 20\,\%.
We have established a tight link between the maximum amplitude frequency $\numax$ and the red-giant radius from an unbiased analysis in the range [7 - 30\,$R_\odot$] which encompasses the red-clump stars. This relation scales as $R\ind{RG} \propto \numax^{-0.48}$.

- From this result, and taking into account the scaling law $\dnumoy \propto \numax^{0.75}$, we have shown that the ratio $\numax / \nuc$ is constant for giants. A similar analysis performed on main-sequence stars and sub-giants reaches the same result: $\numax / \nuc$ is also nearly constant.

- As a by-product, we have shown that scaling laws are slightly but undoubtedly different between giants, sub-giants and dwarfs. For red-giant stars only, the fact that the temperature is nearly a degenerate parameter plays a significant role. As a consequence, global fits encompassing \emph{all} stars with solar-like oscillations may be not precise, since they do not account for the different physical conditions between main-sequence and giant stars.

- The comparison of data from 2 runs pointing in different directions at different galactic latitudes has shown that the stellar properties are similar; the dispersion around the global fits are too weak to be noticed.
The main difference between the 2 runs consists in different stellar populations. The distributions of the asteroseismic parameters are globally similar, except for the location of the red clump.

- We have obtained precise information for the red-clump stars. Statistical asteroseismology makes it possible to identify the expected secondary clump and to measure the distribution of the fundamental parameters of the red-clump stars. We have shown that the relative importance of the two components of the clump is linked to the stellar population. The precise determination of the red-clump parameters will benefit from the asteroseismic analysis and the modeling of individual members of the clump.
\\

These points demonstrate the huge potential of asteroseismology for stellar physics.

\begin{acknowledgements}
This work was supported by the Centre National d'\'Etudes Spatiales (CNES). It is based on observations with CoRoT. The research has made use of the Exo-Dat database, operated at LAM-OAMP, Marseille, France, on behalf of the CoRoT/Exoplanet program. TM acknowledges financial support from Belspo for contract PRODEX-GAIA DPAC.

The work of KB was supported through a postdoctoral fellowship from the `Subside f\'ed\'eral
pour la recherche 2010', Universit\'e de Li\`ege.
SH acknowledges support by the UK Science and Technology Facilities Council. The research leading to these results has received funding from the European Research Council under the European Community's Seventh Framework Programme (FP7/2007--2013)/ERC grant agreement n$^\circ$227224 (PROSPERITY), as well as from the Research Council of K.U.Leuven grant agreement GOA/2008/04.
\end{acknowledgements}

\listofobjects


\begin{thebibliography}
\expandafter\ifx\csname natexlab\endcsname\relax\def\natexlab#1{#1}\fi

\bibitem[{{Alonso} {et~al.}(1999){Alonso}, {Arribas}, \&
  {Mart{\'{\i}}nez-Roger}}]{1999A&AS..140..261A}
{Alonso}, A., {Arribas}, S., \& {Mart{\'{\i}}nez-Roger}, C. 1999, \aaps, 140,
  261

\bibitem[{{Appourchaux}(2004)}]{2004A&A...428.1039A}
{Appourchaux}, T. 2004, \aap, 428, 1039

\bibitem[{{Appourchaux} {et~al.}(2008){Appourchaux}, {Michel}, {Auvergne},
  {Baglin}, {Toutain}, {Baudin}, {Benomar}, {Chaplin}, {Deheuvels}, {Samadi},
  {Verner}, {Boumier}, {Garc{\'{\i}}a}, {Mosser}, {Hulot}, {Ballot}, {Barban},
  {Elsworth}, {Jim{\'e}nez-Reyes}, {Kjeldsen}, {R{\'e}gulo}, \&
  {Roxburgh}}]{2008A&A...488..705A}
{Appourchaux}, T., {Michel}, E., {Auvergne}, M., {et~al.} 2008, \aap, 488, 705

\bibitem[{{Auvergne} {et~al.}(2009){Auvergne}, {Bodin}, {Boisnard}, {Buey},
  {Chaintreuil}, {Epstein}, {Jouret}, {Lam-Trong}, {Levacher}, {Magnan},
  {Perez}, {Plasson}, {Plesseria}, {Peter}, {Steller}, {Tiph{\`e}ne}, {Baglin},
  {Agogu{\'e}}, {Appourchaux}, {Barbet}, {Beaufort}, {Bellenger}, {Berlin},
  {Bernardi}, {Blouin}, {Boumier}, {Bonneau}, {Briet}, {Butler}, {Cautain},
  {Chiavassa}, {Costes}, {Cuvilho}, {Cunha-Parro}, {de Oliveira Fialho},
  {Decaudin}, {Defise}, {Djalal}, {Docclo}, {Drummond}, {Dupuis}, {Exil},
  {Faur{\'e}}, {Gaboriaud}, {Gamet}, {Gavalda}, {Grolleau}, {Gueguen},
  {Guivarc'h}, {Guterman}, {Hasiba}, {Huntzinger}, {Hustaix}, {Imbert},
  {Jeanville}, {Johlander}, {Jorda}, {Journoud}, {Karioty}, {Kerjean},
  {Lafond}, {Lapeyrere}, {Landiech}, {Larqu{\'e}}, {Laudet}, {Le Merrer},
  {Leporati}, {Leruyet}, {Levieuge}, {Llebaria}, {Martin}, {Mazy}, {Mesnager},
  {Michel}, {Moalic}, {Monjoin}, {Naudet}, {Neukirchner}, {Nguyen-Kim},
  {Ollivier}, {Orcesi}, {Ottacher}, {Oulali}, {Parisot}, {Perruchot},
  {Piacentino}, {Pinheiro da Silva}, {Platzer}, {Pontet}, {Pradines},
  {Quentin}, {Rohbeck}, {Rolland}, {Rollenhagen}, {Romagnan}, {Russ}, {Samadi},
  {Schmidt}, {Schwartz}, {Sebbag}, {Smit}, {Sunter}, {Tello}, {Toulouse},
  {Ulmer}, {Vandermarcq}, {Vergnault}, {Wallner}, {Waultier}, \&
  {Zanatta}}]{2009A&A...506..411A}
{Auvergne}, M., {Bodin}, P., {Boisnard}, L., {et~al.} 2009, \aap, 506, 411

\bibitem[{{Ballot} {et~al.}(2010){Ballot}, {Gizon}, {Samadi}, {Vauclair},
  {Appourchaux}, \& {Auvergne}}]{Ballot2010}
{Ballot}, J., {Gizon}, L., {Samadi}, R., {et~al.} 2010, submitted to \aap

\bibitem[{{Barban} {et~al.}(2010){Barban}, {Baudin}, {Mosser}, {Samadi},
  {Goupil}, \& {Michel}}]{Barban2010}
{Barban}, C., {Baudin}, F., {Mosser}, B., {et~al.} 2010, submitted to \aap

\bibitem[{{Barban} {et~al.}(2004){Barban}, {De Ridder}, {Mazumdar}, {Carrier},
  {Eggenberger}, {De Ruyter}, {Vanautgaerden}, {Bouchy}, \&
  {Aerts}}]{2004ESASP.559..113B}
{Barban}, C., {De Ridder}, J., {Mazumdar}, A., {et~al.} 2004, in ESA Special
  Publication, Vol. 559, SOHO 14 Helio- and Asteroseismology: Towards a Golden
  Future, ed. {D.~Danesy}, 113

\bibitem[{{Barban} {et~al.}(2009){Barban}, {Deheuvels}, {Baudin},
  {Appourchaux}, {Auvergne}, {Ballot}, {Boumier}, {Chaplin}, {Garc{\'{\i}}a},
  {Gaulme}, {Michel}, {Mosser}, {R{\'e}gulo}, {Roxburgh}, {Verner}, {Baglin},
  {Catala}, {Samadi}, {Bruntt}, {Elsworth}, \& {Mathur}}]{2009A&A...506...51B}
{Barban}, C., {Deheuvels}, S., {Baudin}, F., {et~al.} 2009, \aap, 506, 51

\bibitem[{{Barban} {et~al.}(2007){Barban}, {Matthews}, {De Ridder}, {Baudin},
  {Kuschnig}, {Mazumdar}, {Samadi}, {Guenther}, {Moffat}, {Rucinski},
  {Sasselov}, {Walker}, \& {Weiss}}]{2007A&A...468.1033B}
{Barban}, C., {Matthews}, J.~M., {De Ridder}, J., {et~al.} 2007, \aap, 468,
  1033

\bibitem[{{Baudin} {et~al.}(2010){Baudin}, {Barban}, {Belkacem}, {Hekker},
  {Morel}, {Samadi}, \& {Goupil}}]{Baudin2010}
{Baudin}, F., {Barban}, C., {Belkacem}, K., {et~al.} 2010, submitted to \aap

\bibitem[{{Bedding} {et~al.}(2010){Bedding}, {Huber}, {Stello}, {Elsworth},
  {Hekker}, {Kallinger}, \& {Mathur}}]{Bedding2010}
{Bedding}, T., {Huber}, D., {Stello}, D., {et~al.} 2010, accepted in \apj

\bibitem[{{Benomar} {et~al.}(2009){Benomar}, {Baudin}, {Campante}, {Chaplin},
  {Garc{\'{\i}}a}, {Gaulme}, {Toutain}, {Verner}, {Appourchaux}, {Ballot},
  {Barban}, {Elsworth}, {Mathur}, {Mosser}, {R{\'e}gulo}, {Roxburgh},
  {Auvergne}, {Baglin}, {Catala}, {Michel}, \& {Samadi}}]{2009A&A...507L..13B}
{Benomar}, O., {Baudin}, F., {Campante}, T.~L., {et~al.} 2009, \aap, 507, L13

\bibitem[{{Brown} {et~al.}(1991){Brown}, {Gilliland}, {Noyes}, \&
  {Ramsey}}]{1991ApJ...368..599B}
{Brown}, T.~M., {Gilliland}, R.~L., {Noyes}, R.~W., \& {Ramsey}, L.~W. 1991,
  \apj, 368, 599

\bibitem[{{Carrier} {et~al.}(2010){Carrier}, {De Ridder}, {Baudin}, {Barban},
  {Hatzes}, {Hekker}, {Kallinger}, {Miglio}, {Montalb{\'a}n}, {Morel}, {Weiss},
  {Auvergne}, {Baglin}, {Catala}, {Michel}, \& {Samadi}}]{2010A&A...509A..73C}
{Carrier}, F., {De Ridder}, J., {Baudin}, F., {et~al.} 2010, \aap, 509, A73+

\bibitem[{{De Ridder} {et~al.}(2009){De Ridder}, {Barban}, {Baudin}, {Carrier},
  {Hatzes}, {Hekker}, {Kallinger}, {Weiss}, {Baglin}, {Auvergne}, {Samadi},
  {Barge}, \& {Deleuil}}]{2009Natur.459..398D}
{De Ridder}, J., {Barban}, C., {Baudin}, F., {et~al.} 2009, \nat, 459, 398

\bibitem[{{Deheuvels} {et~al.}(2010){Deheuvels}, {Bruntt}, {Michel}, {Barban},
  {Verner}, {Regulo}, \& {Mosser}}]{Deheuvels2010}
{Deheuvels}, S., {Bruntt}, H., {Michel}, E., {et~al.} 2010, accepted in \aap

\bibitem[{{Deleuil} {et~al.}(2009){Deleuil}, {Meunier}, {Moutou}, {Surace},
  {Deeg}, {Barbieri}, {Debosscher}, {Almenara}, {Agneray}, {Granet},
  {Guterman}, \& {Hodgkin}}]{2009AJ....138..649D}
{Deleuil}, M., {Meunier}, J.~C., {Moutou}, C., {et~al.} 2009, \aj, 138, 649

\bibitem[{{Dobashi} {et~al.}(2005){Dobashi}, {Uehara}, {Kandori}, {Sakurai},
  {Kaiden}, {Umemoto}, \& {Sato}}]{2005PASJ...57..417D}
{Dobashi}, K., {Uehara}, H., {Kandori}, R., {et~al.} 2005, \pasj, 57, S1

\bibitem[{{Dupret} {et~al.}(2009){Dupret}, {Belkacem}, {Samadi}, {Montalban},
  {Moreira}, {Miglio}, {Godart}, {Ventura}, {Ludwig}, {Grigahc{\`e}ne},
  {Goupil}, {Noels}, \& {Caffau}}]{2009A&A...506...57D}
{Dupret}, M., {Belkacem}, K., {Samadi}, R., {et~al.} 2009, \aap, 506, 57

\bibitem[{{Frandsen} {et~al.}(2002){Frandsen}, {Carrier}, {Aerts}, {Stello},
  {Maas}, {Burnet}, {Bruntt}, {Teixeira}, {de Medeiros}, {Bouchy}, {Kjeldsen},
  {Pijpers}, \& {Christensen-Dalsgaard}}]{2002A&A...394L...5F}
{Frandsen}, S., {Carrier}, F., {Aerts}, C., {et~al.} 2002, \aap, 394, L5

\bibitem[{{Garc{\'{\i}}a} {et~al.}(2009){Garc{\'{\i}}a}, {R{\'e}gulo},
  {Samadi}, {Ballot}, {Barban}, {Benomar}, {Chaplin}, {Gaulme}, {Appourchaux},
  {Mathur}, {Mosser}, {Toutain}, {Verner}, {Auvergne}, {Baglin}, {Baudin},
  {Boumier}, {Bruntt}, {Catala}, {Deheuvels}, {Elsworth}, {Jim{\'e}nez-Reyes},
  {Michel}, {P{\'e}rez Hern{\'a}ndez}, {Roxburgh}, \&
  {Salabert}}]{2009A&A...506...41G}
{Garc{\'{\i}}a}, R.~A., {R{\'e}gulo}, C., {Samadi}, R., {et~al.} 2009, \aap,
  506, 41

\bibitem[{{Gaulme} {et~al.}(2010){Gaulme}, {Deheuvels}, {Mosser}, {Vannier},
  {Guillot}, {Mosser}, {Appourchaux}, \& {Bruntt}}]{Gaulme2010}
{Gaulme}, P., {Deheuvels}, S., {Mosser}, B., {et~al.} 2010, submitted to \aap

\bibitem[{{Girardi}(1999)}]{1999MNRAS.308..818G}
{Girardi}, L. 1999, \mnras, 308, 818

\bibitem[{{Hekker} {et~al.}(2006){Hekker}, {Aerts}, {De Ridder}, \&
  {Carrier}}]{2006A&A...458..931H}
{Hekker}, S., {Aerts}, C., {De Ridder}, J., \& {Carrier}, F. 2006, \aap, 458,
  931

\bibitem[{{Hekker} {et~al.}(2009){Hekker}, {Kallinger}, {Baudin}, {De Ridder},
  {Barban}, {Carrier}, {Hatzes}, {Weiss}, \& {Baglin}}]{2009A&A...506..465H}
{Hekker}, S., {Kallinger}, T., {Baudin}, F., {et~al.} 2009, \aap, 506, 465

\bibitem[{{Huber} {et~al.}(2009){Huber}, {Stello}, {Bedding}, {Chaplin},
  {Arentoft}, {Quirion}, \& {Kjeldsen}}]{2009CoAst.160...74H}
{Huber}, D., {Stello}, D., {Bedding}, T.~R., {et~al.} 2009, Communications in
  Asteroseismology, 160, 74

\bibitem[{{Kallinger} {et~al.}(2008){Kallinger}, {Guenther}, {Matthews},
  {Weiss}, {Huber}, {Kuschnig}, {Moffat}, {Rucinski}, \&
  {Sasselov}}]{2008A&A...478..497K}
{Kallinger}, T., {Guenther}, D.~B., {Matthews}, J.~M., {et~al.} 2008, \aap,
  478, 497

\bibitem[{{Kallinger} {et~al.}(2010){Kallinger}, {Weiss}, {Barban}, {Carrier},
  {De Ridder}, {Hekker}, \& {Samadi}}]{2009ASPC..404..307K}
{Kallinger}, T., {Weiss}, W.~W., {Barban}, C., {et~al.} 2010

\bibitem[{{Kjeldsen} \& {Bedding}(1995)}]{1995A&A...293...87K}
{Kjeldsen}, H. \& {Bedding}, T.~R. 1995, \aap, 293, 87

\bibitem[{{Mathur} {et~al.}(2010{\natexlab{a}}){Mathur}, {Garcia}, {Catala},
  {Appourchaux}, {Auvergne}, \& {Ballot}}]{Mathur2010}
{Mathur}, S., {Garcia}, R., {Catala}, C., {et~al.} 2010{\natexlab{a}},
  submitted to \aap

\bibitem[{{Mathur} {et~al.}(2010{\natexlab{b}}){Mathur}, {Garc{\'{\i}}a},
  {R{\'e}gulo}, {Creevey}, {Ballot}, {Salabert}, {Arentoft}, {Quirion},
  {Chaplin}, \& {Kjeldsen}}]{2010A&A...511A..46M}
{Mathur}, S., {Garc{\'{\i}}a}, R.~A., {R{\'e}gulo}, C., {et~al.}
  2010{\natexlab{b}}, \aap, 511, A46+

\bibitem[{{Mazumdar} {et~al.}(2009){Mazumdar}, {M{\'e}rand}, {Demarque},
  {Kervella}, {Barban}, {Baudin}, {Coud{\'e} Du Foresto}, {Farrington},
  {Goldfinger}, {Goupil}, {Josselin}, {Kuschnig}, {McAlister}, {Matthews},
  {Ridgway}, {Sturmann}, {Sturmann}, {Ten Brummelaar}, \&
  {Turner}}]{2009A&A...503..521M}
{Mazumdar}, A., {M{\'e}rand}, A., {Demarque}, P., {et~al.} 2009, \aap, 503, 521

\bibitem[{{Michel} {et~al.}(2009){Michel}, {Samadi}, {Baudin}, {Barban},
  {Appourchaux}, \& {Auvergne}}]{2009A&A...495..979M}
{Michel}, E., {Samadi}, R., {Baudin}, F., {et~al.} 2009, \aap, 495, 979

\bibitem[{{Miglio} {et~al.}(2009){Miglio}, {Montalb{\'a}n}, {Baudin},
  {Eggenberger}, {Noels}, {Hekker}, {De Ridder}, {Weiss}, \&
  {Baglin}}]{2009A&A...503L..21M}
{Miglio}, A., {Montalb{\'a}n}, J., {Baudin}, F., {et~al.} 2009, \aap, 503, L21

\bibitem[{{Miglio} {et~al.}(2010){Miglio}, {Montalb{\'a}n}, {Carrier}, {Noels},
  \& J.}]{Miglio2010}
{Miglio}, A., {Montalb{\'a}n}, J., {Carrier}, F., {Noels}, A., \& J., D. 2010,
  submitted to \aap

\bibitem[{{Mosser} \& {Appourchaux}(2009)}]{2009A&A...508..877M}
{Mosser}, B. \& {Appourchaux}, T. 2009, \aap, 508, 877

\bibitem[{{Mosser} \& {Appourchaux}(2010)}]{ma10}
{Mosser}, B. \& {Appourchaux}, T. 2010, in New insights into the Sun, ed. {M.
  Cunha, M. Monteiro}, Vol. in press

\bibitem[{{Mosser} {et~al.}(2005){Mosser}, {Bouchy}, {Catala}, {Michel},
  {Samadi}, {Th{\'e}venin}, {Eggenberger}, {Sosnowska}, {Moutou}, \&
  {Baglin}}]{2005A&A...431L..13M}
{Mosser}, B., {Bouchy}, F., {Catala}, C., {et~al.} 2005, \aap, 431, L13

\bibitem[{{Mosser} {et~al.}(2008){Mosser}, {Deheuvels}, {Michel},
  {Th{\'e}venin}, {Dupret}, {Samadi}, {Barban}, \&
  {Goupil}}]{2008A&A...488..635M}
{Mosser}, B., {Deheuvels}, S., {Michel}, E., {et~al.} 2008, \aap, 488, 635

\bibitem[{{Mosser} {et~al.}(2009){Mosser}, {Michel}, {Appourchaux}, {Barban},
  {Baudin}, {Boumier}, {Bruntt}, {Catala}, {Deheuvels}, {Garc{\'{\i}}a},
  {Gaulme}, {Regulo}, {Roxburgh}, {Samadi}, {Verner}, {Auvergne}, {Baglin},
  {Ballot}, {Benomar}, \& {Mathur}}]{2009A&A...506...33M}
{Mosser}, B., {Michel}, E., {Appourchaux}, T., {et~al.} 2009, \aap, 506, 33

\bibitem[{{Rowles} \& {Froebrich}(2009)}]{2009MNRAS.395.1640R}
{Rowles}, J. \& {Froebrich}, D. 2009, \mnras, 395, 1640

\bibitem[{{Roxburgh} \& {Vorontsov}(2006)}]{2006MNRAS.369.1491R}
{Roxburgh}, I.~W. \& {Vorontsov}, S.~V. 2006, \mnras, 369, 1491

\bibitem[{{Samadi} {et~al.}(2007){Samadi}, {Georgobiani}, {Trampedach},
  {Goupil}, {Stein}, \& {Nordlund}}]{2007A&A...463..297S}
{Samadi}, R., {Georgobiani}, D., {Trampedach}, R., {et~al.} 2007, \aap, 463,
  297

\bibitem[{{Stello} {et~al.}(2010){Stello}, {Basu}, {Bruntt}, {Mosser}, \&
  {Stevens}}]{Stello2010}
{Stello}, D., {Basu}, S., {Bruntt}, H., {Mosser}, B., \& {Stevens}, I. 2010,
  accepted in \apj

\bibitem[{{Stello} {et~al.}(2009){Stello}, {Chaplin}, {Basu}, {Elsworth}, \&
  {Bedding}}]{2009MNRAS.400L..80S}
{Stello}, D., {Chaplin}, W.~J., {Basu}, S., {Elsworth}, Y., \& {Bedding}, T.~R.
  2009, \mnras, 400, L80

\bibitem[{{Teixeira} {et~al.}(2009){Teixeira}, {Kjeldsen}, {Bedding}, {Bouchy},
  {Christensen-Dalsgaard}, {Cunha}, {Dall}, {Frandsen}, {Karoff}, {Monteiro},
  \& {Pijpers}}]{2009A&A...494..237T}
{Teixeira}, T.~C., {Kjeldsen}, H., {Bedding}, T.~R., {et~al.} 2009, \aap, 494,
  237

\end{thebibliography}


\end{document}